\documentclass[prl,reprint,a4paper,superscriptaddress,citeautoscript,showpacs,twoside]{revtex4-1}
\usepackage[charter]{mathdesign}
\usepackage{amsmath}
\usepackage[pdftex]{graphicx}
\usepackage{microtype}
\usepackage{bm}\let\vec\bm
\usepackage[svgnames]{xcolor}
\usepackage[colorlinks=True,linkcolor=DarkRed,citecolor=ForestGreen,urlcolor=MediumBlue,pdfstartview=FitH,bookmarks=False,pdfpagemode=UseNone]{hyperref}
\def\sro{Sr$_2$RuO$_4$}
\def\TFL{T_{\mathrm{FL}}}
\def\kB{k_{\mathrm{B}}}

\def\ab{a\!b}
\begin{document}
\title{\boldmath Optical Response of Sr$_2$RuO$_4$ Reveals Universal Fermi-liquid Scaling\\and Quasiparticles Beyond Landau Theory}
\author{D. Stricker}
\affiliation{D{\'e}partement de Physique de la Mati{\`e}re Condens{\'e}e, Universit{\'e} de Gen{\`e}ve, 24 quai Ernest-Ansermet, 1211 Gen{\`e}ve 4, Switzerland}
\author{J. Mravlje}
\affiliation{Jo\v{z}ef Stefan Institute, Jamova 39, 1000 Ljubljana, Slovenia}
\author{C. Berthod}
\affiliation{D{\'e}partement de Physique de la Mati{\`e}re Condens{\'e}e, Universit{\'e} de Gen{\`e}ve, 24 quai Ernest-Ansermet, 1211 Gen{\`e}ve 4, Switzerland}
\author{R. Fittipaldi}
\affiliation{CNR-SPIN, and Dipartimento di Fisica ``E. R. Caianiello'', Universita di Salerno, I-84084 Fisciano (Salerno) Italy}
\author{A. Vecchione}
\affiliation{CNR-SPIN, and Dipartimento di Fisica ``E. R. Caianiello'', Universita di Salerno, I-84084 Fisciano (Salerno) Italy}
\author{A. Georges}
\affiliation{Coll{\`e}ge de France, 11 place Marcelin Berthelot, 75005 Paris, France}
\affiliation{Centre de Physique Th{\'e}orique, {\'E}cole Polytechnique, CNRS, 91128 Palaiseau, France}
\affiliation{D{\'e}partement de Physique de la Mati{\`e}re Condens{\'e}e, Universit{\'e} de Gen{\`e}ve, 24 quai Ernest-Ansermet, 1211 Gen{\`e}ve 4, Switzerland}
\author{D. van der Marel}
\affiliation{D{\'e}partement de Physique de la Mati{\`e}re Condens{\'e}e, Universit{\'e} de Gen{\`e}ve, 24 quai Ernest-Ansermet, 1211 Gen{\`e}ve 4, Switzerland}
\begin{abstract}
We report optical measurements demonstrating that the low-energy relaxation rate ($1/\tau$) of the conduction electrons 
in Sr$_2$RuO$_4$ obeys scaling relations for its frequency ($\omega$) and temperature ($T$) dependence in accordance 
with Fermi-liquid theory. 
In the thermal relaxation regime, $1/\tau\propto (\hbar\omega)^2 + (p\pi\kB T)^2$ with $p=2$, and $\omega/T$ scaling applies.
Many-body electronic structure calculations using dynamical mean-field theory confirm the low-energy Fermi-liquid scaling,  
and provide quantitative understanding of the deviations from Fermi-liquid behavior at higher energy and temperature.  
The excess optical spectral weight in this regime provides evidence for strongly dispersing ``resilient'' quasiparticle excitations above the Fermi energy. 
\end{abstract}
\maketitle
Liquids of interacting fermions yield a number of different emergent states of quantum matter. 
The strong correlations between their constituent particles pose a formidable theoretical challenge.
It is therefore remarkable that a simple description of low-energy excitations of fermionic quantum liquids could be established early on by Landau \cite{Landau-1956}, in terms of a dilute gas of ``quasiparticles'' with a renormalized effective mass, of which $^3$He is the best documented case\cite{baym_pethick_book,Leggett-2004}.

Breakdown of the quasiparticle concept can be observed in the transport of metals tuned 
onto a quantum phase transition, but Fermi-liquid (FL) behavior is retrieved away from the quantum-critical 
region 
\cite{sachdev_book,Cubrovic-2009}.
The relevance of FL theory to electrons in solids is documented by a number of materials, such as transition metals \cite{Rice-1968}, heavy-fermion compounds \cite{Kadowaki-1986}, and doped semiconductors \cite{vanderMarel-2011}. 
Among transition-metal oxides, \sro{} is a remarkable example which has been heralded as the solid-state analogue of  $^3$He \cite{Mackenzie-2003} 
for at least three reasons: remarkably large and clean monocrystalline samples can be prepared, transport properties display   
low-temperature FL characteristics \cite{Hussey-1998}, and there is evidence for $p$-wave symmetry 
of its superconducting phase \cite{Kallin-2012}, as in superfluid $^3$He. 

FL theory makes a specific prediction for the universal energy and temperature dependence of the inelastic lifetime of quasiparticles: Because 
of phase-space constraints imposed by the Pauli principle as well as momentum and energy conservation, it 
diverges as $1/\omega^2$ or $1/T^2$ \cite{Landau-1956,Cubrovic-2009}. 
More precisely, the inelastic optical relaxation rate is predicted to vanish according to the scaling law $1/\tau\propto (\hbar\omega)^2 + (p\pi\kB T)^2$, 
with $p=2$ \cite{Gurzhi-1959, Chubukov-2012, Berthod-2013}. This leads to universal $\omega/T$ scaling of the optical conductivity $\sigma(\omega)$
in the thermal  regime $\hbar\omega\sim\kB T$ \cite{Berthod-2013}.
Surprisingly however, despite almost 60 years of research on Fermi liquids, this universal behavior of the optical response, and 
especially the specific statistical factor $p=2$ relating the energy and temperature dependence have not yet been established experimentally \cite{Basov-2011,Chubukov-2012,Berthod-2013,Nagel-2012,Mirzaei-2013}.  

Here, we report optical measurements of \sro{} with 0.1~meV resolution \cite{PhysRevLett.94.107003,Ingle-2005} which reveal this universal FL scaling law \footnote{Note that the resolution in the ARPES studies of Refs.~\onlinecite{PhysRevLett.94.107003} and \cite{Ingle-2005} was 25 and 14 meV, respectively.}. 
We establish experimentally the universal value ${p~=~2}$ and demonstrate remarkable agreement between the experimental data and 
the theoretically derived scaling functions in the FL regime. 
Importantly, the identification of the precise FL response also enables us to characterize the deviations from FL theory.
The manifestation of these deviations in our data is an excess spectral weight above $0.1$~eV. 
We show that this is an optical fingerprint of the abrupt increase in dispersion of  ``resilient'' quasiparticle excitations. This confirms the recent prediction \cite{Deng-2013} on the basis of dynamical mean-field theory (DMFT), that well-identified peaks in the spectral function persist far above the asymptotic low-energy and low-temperature Landau FL regime where the relaxation rate has a strict $\omega^2$ dependence.
In this Letter, we perform realistic DMFT calculations for Sr$_2$RuO$_4$   which yield excellent agreement with the measured optical spectra. 
There are three bands at $E_F$: a two-dimensional one ($\gamma$) of $d_{xy}$ character and two quasi-1D bands ($\alpha$ and $\beta$), respectively\cite{Mackenzie-2003}.
We show that the deviations from FL behavior in the optical spectra are caused by resilient quasiparticle (QP) excitations associated with unoccupied states, which are inaccessible in usual photoemission experiments.  

\begin{figure}[tb]
\centerline{\includegraphics[width=.9\columnwidth]{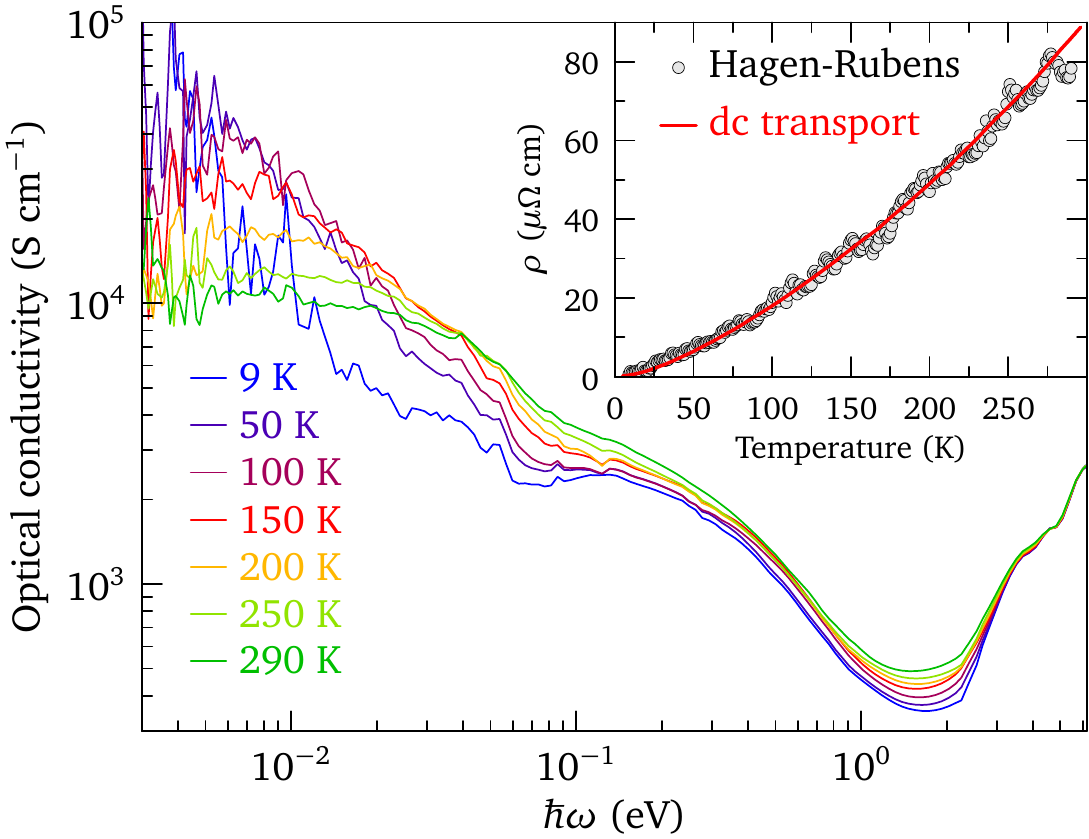}}
\caption{\label{Fig:OpticalConductivity}
Real part of the optical conductivity of \sro{} for selected temperatures between 9 and 290~K. 
Inset: Zero-frequency resistivity determined by the Hagen-Rubens fit of the reflectivity (open circles) and four-terminal dc resistivity of the same crystal (solid red curve) multiplied with a factor 0.84, which is within the range of geometrical factors due to sample shape and contact layout.}
\end{figure}

Optical spectroscopy is a powerful probe of---among other things---the subtle low-energy behavior of electron liquids. Earlier optical studies of \sro{} have reported that the in-plane low-energy spectral weight is about 100 times larger than the one along the $c$ axis, with an onset below 25~K of a $T^2$ relaxation rate \cite{Katsufuji-1996, Hildebrand-2001, Pucher-2003}.  
The lowest-lying interband transitions, located above 1~eV, have been previously identified as $d$-$d$ transitions \cite{Lee-2002}. In the optical conductivity of Sr$_2$RuO$_4$ this is revealed as a peak at 1.7~eV (\cite{SM}, Table~I). In the range displayed in Fig.~\ref{Fig:OpticalConductivity}, $\sigma(\omega)$ is entirely due to the free carrier response. The dynamical character of the inelastic scattering can be captured by a frequency-dependent memory function $M(\omega)$ as described in Ref.~\onlinecite{PhysRevB.6.1226} so that
	\begin{equation}\label{Eq:SigmaMemory}
		\sigma(\omega)=\frac{i\epsilon_0\omega_p^2}{\omega+M(\omega)}.
	\end{equation}
\noindent On the other hand, an intraband optical absorption process excites electron-hole {\it pairs}, with the consequence 
that the optical relaxation rate $M_2(\omega)$ is proportional to $(\hbar\omega)^2+(p\pi\kB T)^2$ with the value $p=2$ \cite{Gurzhi-1959, Chubukov-2012, Berthod-2013}.
The optical conductivity is then characterized by a narrow Lorentzian-like zero-frequency mode (Drude peak), 
followed by a (non-Drude) ``foot'' at $\hbar\omega\approx 2\pi \kB T$ \cite{Berthod-2013}. 
Hence, the signature of FL theory and of the frequency dependence of $1/\tau$ is actually a 
{\it deviation from Drude's form} (corresponding to a constant $\tau$).
Several recent optical studies have reported $\omega^2$ and $T^2$ for $M_2(\omega)$ in a number of different materials.
However, in neither of these cases does the coefficient $p$ match the prediction $p=2$: $p\sim 1$ in URu$_2$Si$_2$ \cite{Nagel-2012}, $p\sim 2.4$ in the organic material BEDT-TTF \cite{Dressel-2011}, and $p\sim 1.5$ in underdoped HgBa$_2$CuO$_{4+\delta}$ \cite{Mirzaei-2013}. One possible scenario that has been proposed to explain this discrepancy is the presence of magnetic impurities \cite{Chubukov-2012}.
We decided instead to look at the $4d$ correlated material \sro which can be synthesized in very pure form, with well-established $T^2$ resistivity below 25~K \cite{Mackenzie-2003}.

The \sro{} crystal employed for this work was grown by sing the travelling floating zone technique \cite{Udagawa-2005}. The quality of the crystal was confirmed by different techniques \cite{Fittipaldi-2005} with a superconducting transition at 1.4~K. 
The $ab$-plane crystal surface of $5.1 \times 3.6$~mm$^2$ was micropolished and cleaned prior to transferring the sample to the UHV cryostats for optical spectroscopy. 
Near-normal reflection reflectivity spectra in the range from 2~meV to 3~eV were collected 
between 290 and 9~K at a cooling rate of 1~K per minute. 
The optical conductivity obtained by Kramers-Kronig analysis is shown in Fig.~\ref{Fig:OpticalConductivity} for a few selected temperatures. 
The close match of the dc resistivity and $\rho(T)=\mbox{lim}_{\omega\rightarrow 0} 1/\sigma_1(\omega,T)$ (inset in Fig.~\ref{Fig:OpticalConductivity}) provides evidence that the low-frequency optical data are accurate at all temperatures.

\begin{figure}[b]
\centerline{\includegraphics[width=\columnwidth]{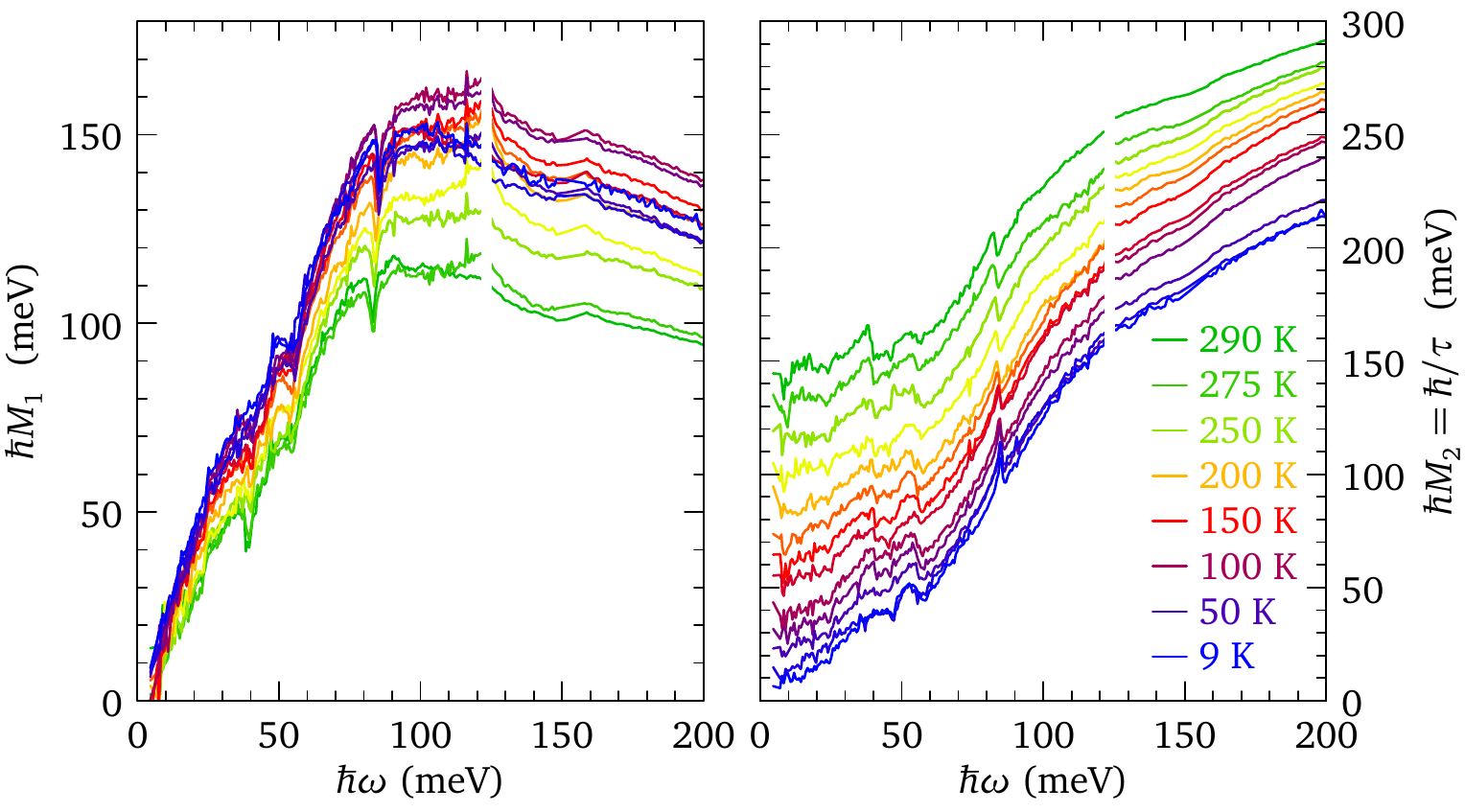}}
\caption{\label{Fig:MemoryFunction}
Real part (left) and imaginary part (right) of the \sro{} memory function for selected temperatures between 9 and 290~K. A white space is introduced near 130 meV where data sets from different detectors were linked.}
\end{figure}

\begin{figure}[tb]
\centerline{\includegraphics[width=.9\columnwidth]{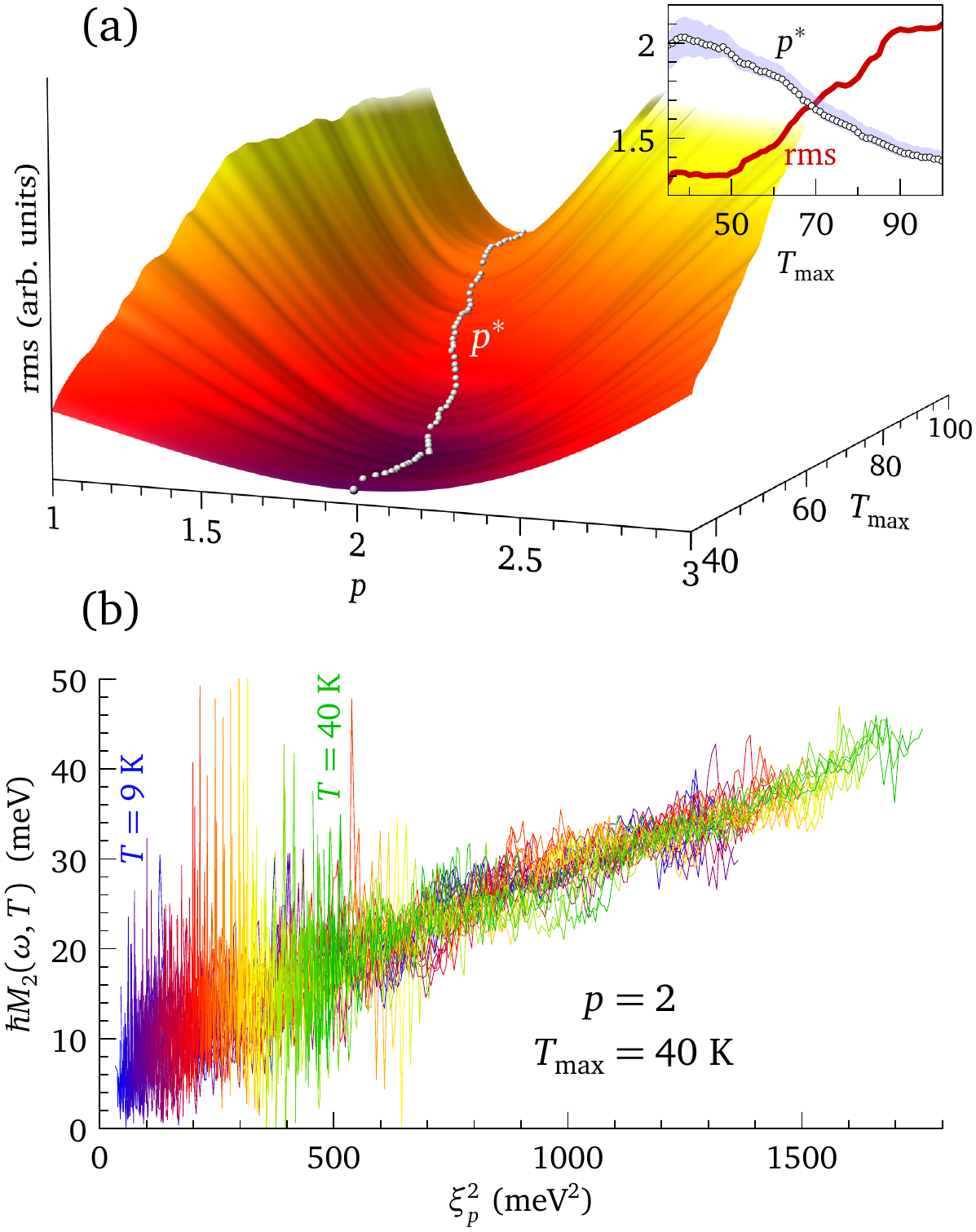}}
\caption{\label{Fig:ScalingRelaxationRate}
(a) Root-mean square deviation of the relaxation rate $M_2(\omega,T)$ from a linear dependence in $\xi_p^2$, for $\hbar\omega\leq 36~\mathrm{meV}$ and $T\leq T_{\max}$, as a function of $p$ and $T_{\max}$. 
The inset shows the value $p^*$ and the rms at the minimum versus $T_{\max}$. 
A value $p^*=2$ is found below $T_{\max}\sim40$~K. The shaded region shows how $p^*$ changes if the frequency range is varied by $\pm5$~meV. (b) Collapse of the relaxation rate data for $T\leq 40$~K. 
}
\end{figure}

\begin{figure}[b]
\centerline{\includegraphics[width=\columnwidth]{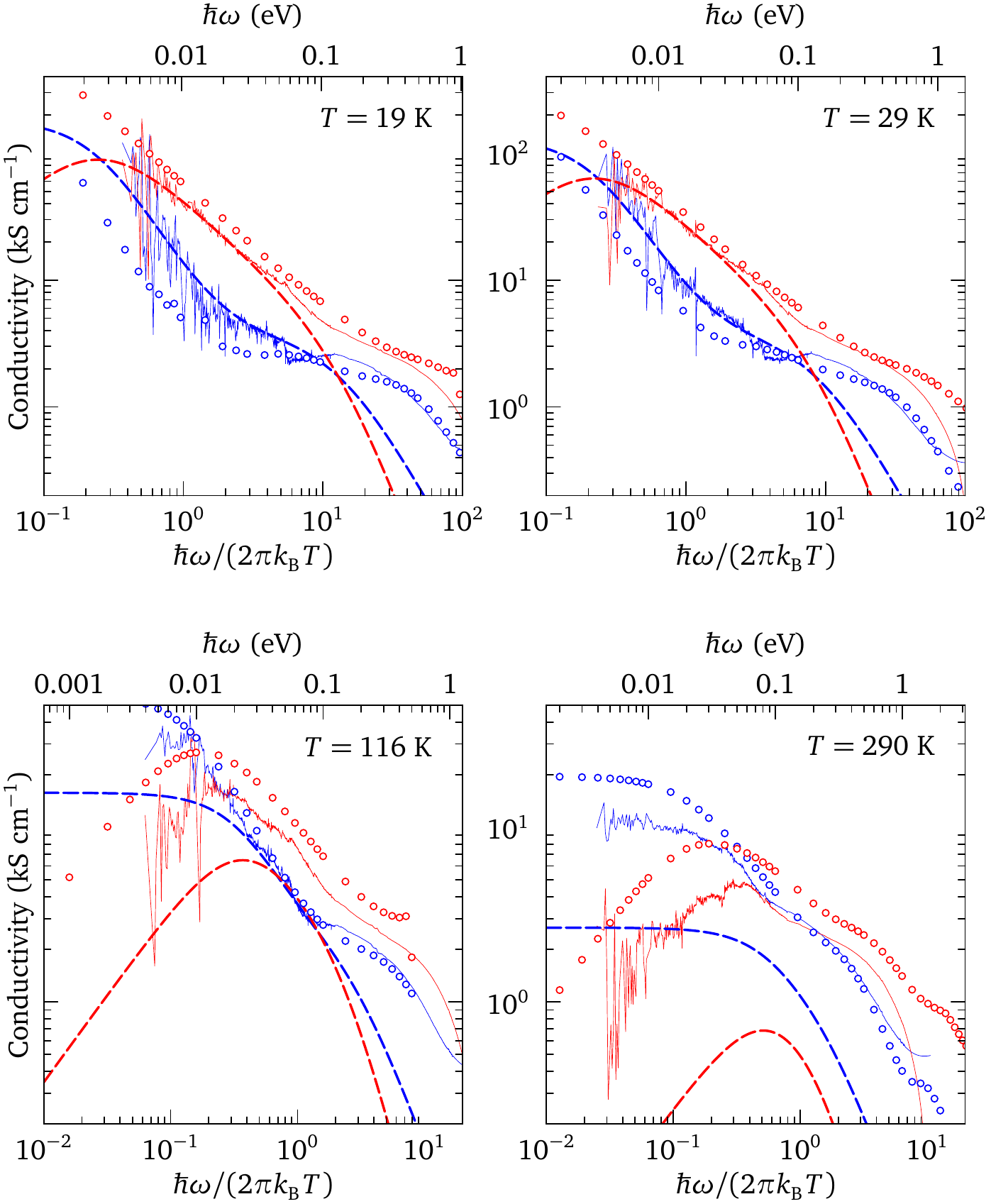}}
\caption{\label{Fig:loglog}
Comparison of the experimental optical conductivity (solid lines), Fermi-liquid model (dashed lines), and DFT+DMFT calculation (circles) at selected temperatures.
The real and imaginary parts of the conductivity are shown in blue and red, respectively. 
The three parameters of the Fermi-liquid model, including impurity scattering, are fit to the experimental data in the range $\hbar\omega\leq36~\mathrm{meV}$ and $T\leq 40$~K.
The DFT+DMFT calculation ignores impurity scattering.}
\end{figure}

The optical conductivity displayed in Fig.~\ref{Fig:OpticalConductivity} is dominated by the peak centered at zero frequency, corresponding to the optical response of the free charge carriers. Upon lowering the temperature from 290 to 9~K this peak becomes extremely narrow, and its maximum at  $\omega=0$ increases by 2 orders of magnitude. The weak features at 40, 57, and 85 meV correspond to optical phonons.
The standard Drude model assumes a frequency-independent relaxation rate.
The frequency dependence of $M_2(\omega)$ shown in Fig.~\ref{Fig:MemoryFunction}(b) is therefore manifestly non-Drude like, and signals the presence of a dynamical component in the quasiparticle self-energy.
Moreover, below 0.1~eV, $M_2(\omega)$ has a positive curvature for all temperatures corresponding to $\omega^{\eta}$ with $\eta\approx 2$ (\cite{SM}, Sec.~III), and $M_1(\omega)$ has a linear frequency dependence that is only weakly changing with temperature.
This is the expected behavior in a Fermi liquid. The low-frequency mass enhancement factor $m^*(\omega)/m=1+M_1(\omega)/\omega$ varies from 3.3 at 9~K to 2.3 at 290~K. The $m^*(\omega)/m$ curves (\cite{SM}, Sec.~II) fall slightly below the one of a previous room temperature study \cite{Lee-2006}.
In order to reveal the signature of Fermi-liquid behavior, we searched for the presence of a universal scaling of the form
$M_2(\omega,T)\propto\xi_p^2\equiv(\hbar\omega)^2+(p\pi\kB T)^2$ 
in the data, by plotting $M_2(\omega,T)$ parametrically as a function of $\xi_p^2$ for different choices of $p$ and calculating the root-mean square (rms) deviation of this plot from a straight line.
The frequency range used in this analysis was limited to $\hbar\omega\leq 36~\mathrm{meV}$, and the largest temperature considered, $T_{\max}$, was allowed to vary down to $T_{\max}= 35 $~K, below which the fitted temperature range becomes too small to produce reliable output.
The result of the scaling collapse for $p=2$ and $T\leq T_{\max}=40$~K is displayed in Fig.~\ref{Fig:ScalingRelaxationRate}. 
The rms minimum for each $T_{\max}$ defines $p^*$, shown as a function of $T_{\max}$ in the inset.  
When the range $T_{\max}$ is varied from 100 to 35~K we observe a flow from $p=1.5$ towards the plateau value $p=2$, which is approached for $T_{\max}\leq50$.
This confirms the expectation of a flow towards universal Fermi-liquid behavior for $T\rightarrow 0$, for which we expect a collapse of all data on a universal function
of $\xi_p$ with $p=2$.
A similar analysis conducted on the raw reflectivity data leads to the same conclusion that $p=2$(\cite{SM}, Sec.~III).
A direct confirmation of FL behavior is found in the optical conductivity curves (Fig.~\ref{Fig:loglog}).
They exhibit a characteristic non-Drude feature in perfect agreement with the universal FL response.
This feature is an increase of conductivity with respect to the low-frequency Drude response around the thermal frequency $\hbar\omega=2\pi\kB T$, appearing most clearly as a shoulder in a log-log plot.
The universal FL response, including impurity scattering, has been parametrized by only three temperature-independent parameters \cite{Berthod-2013}.
This three-parameter model can reproduce the low-frequency optical conductivity data of \sro{} in the whole temperature range below 40~K, where $p=2$, as illustrated in Fig.~\ref{Fig:loglog}.

For energies above 0.1~eV and/or temperatures above 40~K, the measured conductivities clearly depart 
from the reference FL (Fig.~\ref{Fig:loglog}). 
All deviations go in the direction of a larger conductivity (both real and imaginary parts), in particular, in the 0.1--0.5~eV energy range.
In order to understand the origin of this increased conductivity, we have calculated 
the optical spectra within an \textit{ab initio} framework that combines density-functional theory (DFT) 
with the many-body DMFT \cite{georges_rmp_1996}, 
as described in \cite{Aichhorn-2009,TRIQS} and applied to \sro{} in \cite{Mravlje-2011}.
The bare dispersions and velocities are obtained from DFT for the three $t_{2g}$ bands, 
and the local (momentum-independent) DMFT self-energies for each orbital are calculated by
using the same interaction parameters as in previous works (\cite{SM}, Sec.~IV).
The theoretical results are presented in Fig.~\ref{Fig:loglog} as circles. 
The overall shapes of experimental data and theoretical DMFT results match closely, and 
satisfactory agreement is also found for absolute values (note that the comparison in 
Fig~\ref{Fig:loglog} involves no scale adjustment).  
At low frequency and temperature, the \textit{ab initio} calculations show a Drude peak and a thermal shoulder, 
in excellent agreement with the experimental data and with the FL model.
The differences at the lowest frequencies can be attributed to impurity scattering, included in the FL model but not in 
the DMFT calculation, in order to keep the latter parameter-free.
More interestingly, above 0.1~eV the theory deviates from the FL model in precisely the same manner as the experiment does.
At higher temperatures, while the predictions extrapolated from the low-temperature FL severely underestimate the conductivity, 
the agreement between DFT+DMFT and experiments remains excellent in the 0.1--0.5~eV range.
The calculated imaginary part of the optical conductivity is systematically somewhat lower than the experimental data. 
The difference increases with temperature, and is most clear for 290 K. 
Electron-phonon interactions in fact cause additional mass-enhancement, which leads to a suppression 
of both $\sigma_1$ and $\sigma_2$ for $\omega\rightarrow 0$ and an increase of optical conductivity in the phonon energy range. 
This effect is not included in the DMFT calculations and may explain the remaining differences with experimental data. 

We have carried out a series of numerical experiments in order to elucidate the origin of the non-FL excess of 
optical spectral weight in the $0.1$--$0.5$~eV frequency range (\cite{SM}, Sec.~IV).
First, we eliminated band-structure effects as a possible cause.
The band structure enters the optical conductivity via a transport function $\Phi(\varepsilon)$, 
proportional to the average of the squared velocities at a given energy.
This function is smooth, unlike the density of states which diverges at the van Hove singularity of the $xy$ band. 
Indeed, we have verified that the replacement of $\Phi(\varepsilon)$ by its Fermi-surface value $\Phi(0)$ causes 
no significant change in the theoretical curves of Fig.~\ref{Fig:loglog}.
The excess spectral weight is therefore due to electronic correlations 
and must be linked to a structure in the single-particle self-energies.

The imaginary part of the self-energies follows the FL parabolic dependence at low-energy 
but starts to deviate already well below 0.1~eV in the direction of a weaker scattering.
In particular, between $+0.2$ and $+0.4$~eV, still in the domain of intraband transitions, 
the scattering rate goes through a maximum and decreases slightly at higher energy.
A similar phenomenon with a saturation of the scattering rate was observed in the single-band Hubbard model 
and was shown to give rise to resilient QPs \cite{Deng-2013}.
In \sro{}, the signature of resilient QPs is even more striking: 
It is signaled by a {\it drop} of the scattering rate for empty states above $\sim0.35$~eV.
Consistently with Kramers-Kronig relations, this drop implies a sharp minimum in the real part of the self-energy which is 
found in the energy range 0.1--0.15~eV. 
As a consequence,  QPs above this energy scale have velocities \emph{larger} than the bare velocities.
In the theoretical spectral function, these appear as peaks which are broader than the low-energy Landau QP peaks and 
have a very steep dispersion in the range 0.2--0.4 eV, leading to an inverted waterfall-like structure.

These resilient QPs with large velocities are the source of excess spectral weight and deviation from FL 
behavior above 0.1~eV. 
Indeed, optical spectroscopy is sensitive to these excitations above the Fermi level since it probes transitions 
between occupied and unoccupied states. 
The abrupt increase of QP velocities 
predicted by DFT+DMFT results in a maximum in the 
real part of the calculated memory function $M_1(\omega)$ in the range 0.1--0.2~eV, 
hence providing an explanation for the corresponding feature found experimentally 
(Fig.~\ref{Fig:MemoryFunction}). 
We also note that more subtle changes in QP dispersions (kinks) at $\sim 30$~meV, previously found 
in both angle-resolved photoemission spectroscopy (ARPES) \cite{Ingle-2005,Iwasawa-2013} and DMFT \cite{Mravlje-2011}, are also visible in 
$M_1$ and $M_2$ at lower energy but do not change the frequency dependence of the optical conductivity 
so strikingly. 

The resilient QP excitations above the Fermi level predicted by our calculations and 
leading to the sharp feature in $M_1$ are not directly accessible to conventional ARPES, which probes only occupied states. 
Recently, two-photon ARPES has been shown to provide energy and momentum-resolved information on 
unoccupied states \cite{sobota_ARPES_2photons_prl_2013}, and we propose that our theoretical results 
(\cite{SM}, Fig.~SM14) could be put to the test in the future by using this technique for \sro. 

In summary, we have performed reflectance and ellipsometry measurements for a \sro{} single crystal in a wide range of frequencies and temperatures 
and observed for the first time the universal optical signatures of the Landau quasiparticles in a FL.
The low-energy optical relaxation rate obeys $(\hbar\omega)^2+(p\pi\kB T)^2$ scaling with $p=2$, and the optical conductivity exhibits a pronounced non-Drude foot at the thermal frequency $\hbar\omega=2\pi\kB T$.
The identification of a low-energy FL regime provides a reference to characterize without ambiguity the deviations from FL theory.
In \sro{}, the most significant deviation at low temperature is an increase of conductivity developing above $0.1$~eV.
With the help of DFT+DMFT calculations, we ascribed the extra spectral weight to resilient quasiparticle 
excitations above the Fermi level, i.e., relatively broad but still 
strongly dispersing particlelike excitations with a  lifetime differing from the Landau low-energy form.
This work demonstrates that optical spectroscopy is a powerful tool to diagnose non-FL behavior, with the provision that the proper FL behavior is taken as the  ``placebo'' reference, instead of the Drude law that is often used for that purpose.

\acknowledgments
We thank F. Baumberger, Y. Maeno, and Z.-X. Shen for stimulating discussions, J. Jacimovic and E. Giannini for assistance with the resistivity experiments, and M. Brandt for technical assistance. This work was supported by the Swiss National Science Foundation (SNSF) through Grants No. 200020-140761 and No. 200021-146586, by the Slovenian research agency program P1-0044, by FP7/2007-2013 through grant No. 264098-MAMA, and by the ERC through Grant No. 319-286 (QMAC). 
Computing time was provided by IDRIS-GENCI and the Swiss CSCS under Project No. S404.  

%%%%%%%%%%%%%%
%\bibliography{MS_bib}%
%%%%%%%%%%%%%%

\onecolumngrid
\newpage
\begin{center}

{\large\textbf{\boldmath
Supplemental Material\\ [0.5em] {\small to} \\ [0.5em]
Optical Response of Sr$_2$RuO$_4$ Reveals Universal Fermi-liquid Scaling\\[0.4em]
and Quasiparticles Beyond Landau Theory}}\\[1.5em]

D. Stricker,$^1$ J. Mravlje,$^2$ C. Berthod,$^1$ R. Fittipaldi,$^3$ A.
Vecchione,$^3$ A. Georges,$^{4,5,1}$ and D. van der Marel$^1$\\[0.5em]

\textit{\small
$^1$D{\'e}partement de Physique de la Mati{\`e}re Condens{\'e}e, Universit{\'e}
de Gen{\`e}ve, 24 quai Ernest-Ansermet, 1211 Gen{\`e}ve 4, Switzerland\\
$^2$Jo\v{z}ef Stefan Institute, Jamova 39, 1000 Ljubljana, Slovenia\\
$^3$CNR-SPIN, and Dipartimento di Fisica ``E. R. Caianiello'', Universita di
Salerno, I-84084 Fisciano (Salerno), Italy\\
$^4$Coll{\`e}ge de France, 11 place Marcelin Berthelot, 75005 Paris, France\\
$^5$Centre de Physique Th{\'e}orique, {\'E}cole Polytechnique, CNRS, 91128
Palaiseau, France
}

\vspace{2em}
\end{center}

\twocolumngrid
\setcounter{figure}{0}
\renewcommand{\thefigure}{SM\arabic{figure}}

\section{Optical spectroscopy}
\label{SM-I}
The measurement of the in-plane optical response was carried out with two FTIR
spectrometers and a Woolam VASE ellipsometer. The far-infrared to near-infrared
reflectivity at near-normal incidence was measured from 1.9~meV to 1.1~eV 
(15 to 9000 cm$^{-1}$) using a Bruker 113 FTIR spectrometer. A Bruker 66 upgraded 
with a high-frequency module extended this range to 3~eV. The ellipsometry
measurements were performed in a Woolam VASE ellipsometer, completing and
extending the data up to the UV (0.47--6.2~eV). The two spectroscopic techniques
thus overlap from 0.47 to 3~eV. Light polarized along the $\ab$ plane
(p-polarization) was used in order to suppress the $c$-axis features, as
described in Ref.~\onlinecite{van_heumen_optical_2007}.

We used Sr$_2$RuO$_4$ crystals containing less than 2 volume percent intercalated layers of Sr$_3$Ru$_2$O$_7$ of three unit cell thickness on average. The crystals show no detectable trace of the 3K phase.
The crystal was cleaved along the $\ab$ plane, polished to achieve a perfectly
reflective surface, and glued onto a copper sample holder using silver paste.
The device was mounted in a high vacuum and high stability home-made cryostat
($\sim 10^{-8}$~mbar) including an \textit{in-situ} gold/silver evaporator to
cover the sample with a reference layer. Temperature sweeps were conducted
between 9~K and 290~K at a speed of 1~K per minute, and one reflectivity
spectrum was collected every Kelvin. The long-term drift of the light sources
and detectors was calibrated using a high stability flipping mirror placed in
front of the cryostat window. The mirror was flipped up and down every 20
minutes, providing a continuous monitoring of the signal drift. The remaining
uncorrected drift was analyzed during the warmup process, which leads to a
thermal hysteresis effect below the noise level in the far IR, below 0.2\% in
the mid IR, and below 0.4\% in the near IR and above.

\begin{figure}
\includegraphics[width=0.85\columnwidth]{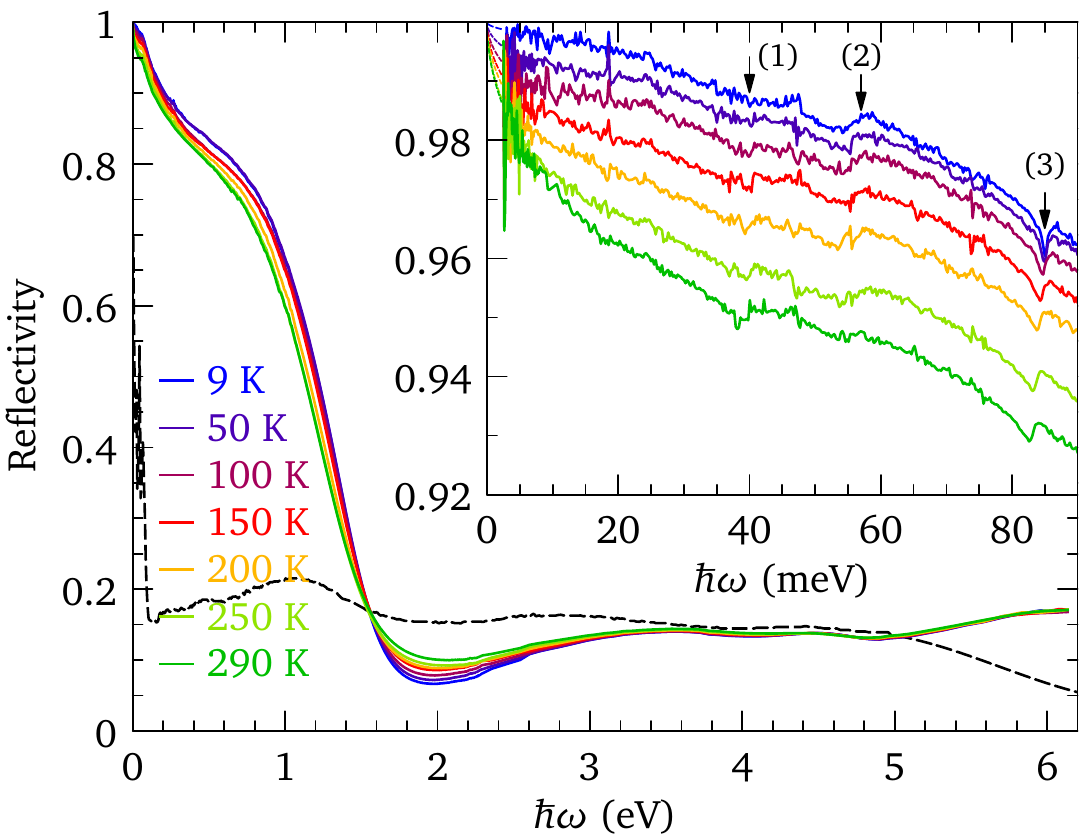}
\caption{\label{FigSup:Reflectivity}
Optical reflectivity of \sro{} along the $\ab$ plane (solid lines) and $c$ axis
(dashed line, reproduced from Ref.~\onlinecite{katsufuji_-plane_1996}) at
selected temperatures. The inset shows the low-frequency region, with the
Hagen-Rubens extrapolation to zero frequency, Eq.~(\ref{EqSup:HagenRubens}),
shown as dashed lines. The 84~meV feature (3) is a longitudinal in-plane
bond-stretching phonon mode of the RuO$_2$ layer (zone center double degenerated
E$_u$). The slight softening and broadening upon heating indicate a decrease of
screening. The structures (1) at 40~meV and (2) at 57~meV most likely also
correspond to phonons.
}
\end{figure}

The calibrated $\ab$-plane reflectivity is obtained as
	\begin{equation}\label{EqSup:Rcorr}
		R(\omega) = \frac{I_{\mathrm{sample}}(\omega)}
		{I_{\mathrm{reference}}(\omega)}
		\frac{I_{\mathrm{reference-mirror}}(\omega)}
		{I_{\mathrm{sample-mirror}}(\omega)},
	\end{equation}
where $I_{\mathrm{sample-mirror}}(\omega)$ and
$I_{\mathrm{reference-mirror}}(\omega)$ are the intensities from the mirror,
measured together with the sample and reference, respectively. The absolute
reflectivity $R(\omega)$ is shown for selected temperatures in
Fig.~\ref{FigSup:Reflectivity}. The figure also shows the room-temperature
$c$-axis reflectivity reproduced from Ref.~\onlinecite{katsufuji_-plane_1996}.

\begin{figure}
\includegraphics[width=0.85\columnwidth]{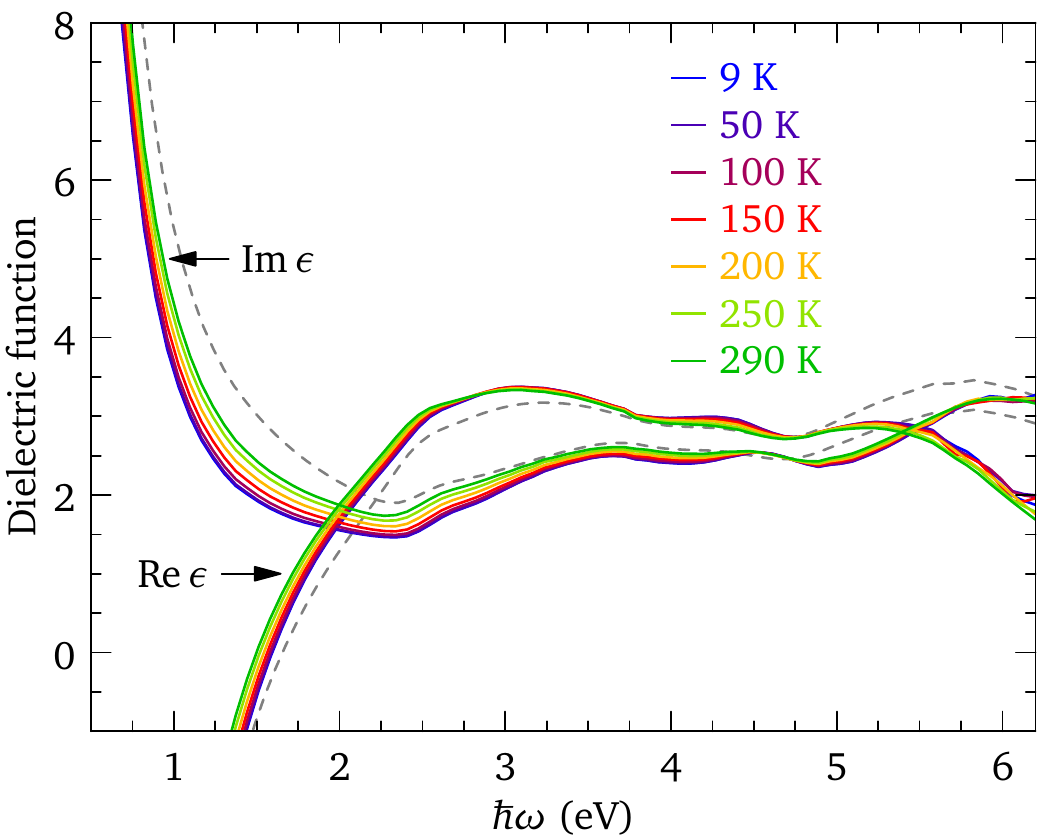}
\caption{\label{FigSup:Sup2pseudoeps}
Real and imaginary parts of the dielectric function $\epsilon$
extracted from Eq.~(\ref{EqSup:pseudoeps}). The dashed lines
show the complex pseudo dielectric function $\tilde{\epsilon}$ at 290~K.
}
\end{figure}

The ellipsometry data is used to extend the frequency range up to 6.2~eV. The
angle of incidence $\theta$ of the light was 70$^{\circ}$ measured from the
surface normal. Owing to the anisotropy of \sro, the measured complex dielectric
function $\tilde{\epsilon}(\omega)$ (the so-called pseudo-dielectric function) is a mix of the
$\ab$-plane and $c$-axis responses. The $\ab$-plane dielectric function
$\epsilon(\omega)=\epsilon_1(\omega)+i\epsilon_2(\omega)$ was extracted from $\tilde{\epsilon}(\omega)$ by inverting the Fresnel
equations, using the complex $c$-axis dielectric function $\epsilon_c(\omega)$ deduced from the
$c$-axis reflectivity of Ref.~\onlinecite{katsufuji_-plane_1996}:
	\begin{multline}\label{EqSup:pseudoeps}
		\tilde{\epsilon} = \sin^2\!\theta+\sin^4\!\theta\\
		\times\left(\frac{\sqrt{\epsilon-\sin^2\!\theta}
		\sqrt{\epsilon^{\phantom 2}}-
		\sqrt{1-\epsilon_c^{-1}\sin^2\!\theta}}{\sqrt{\epsilon-\sin^2\!\theta}
		\sqrt{1-\epsilon_c^{-1}\sin^2\!\theta}
		-\sqrt{\epsilon^{\phantom 2}}\cos^2\!\theta}\right)^2.
	\end{multline}
The resulting $\epsilon(\omega)$ is shown in Fig.~\ref{FigSup:Sup2pseudoeps}.
Following Ref.~\onlinecite{van_heumen_optical_2007}, the consistency of the
procedure for extracting $\epsilon(\omega)$ has been checked by repeating the
ellipsometry measurements at several angles of incidence.
A Kramers-Kronig transform of the reflectivity, using the ellipsometry data to
anchor the phase, provides the complex optical conductivity $\sigma(\omega)=\sigma_1(\omega)+i\sigma_2(\omega)$. 
%The relative accuracy of the low-frequency  conductivity as a function of temperature can be assessed by a comparison with dc transport measurements, after extrapolating $\sigma(\omega)$ to zero frequency. 
The dielectric function and the conductivity are related by $\epsilon(\omega)=1+i\sigma(\omega)/(\epsilon_0\omega)$. The reflection coefficient is given by 
\begin{equation}\label{Eq:R}
R(\omega)=\left|\frac{1-\sqrt{\epsilon(\omega)}}{1+\sqrt{\epsilon(\omega)}}\right|^2 \approx 1-4\mathrm{Re}\,\sqrt{\frac{1}{\epsilon(\omega)}}.
\end{equation} 
The right-hand side of this expression applies to the infrared frequency range, where $|\epsilon(\omega)|\gg 1$. Furthermore for $\omega\ll 1/\tau$ the conductivity can be approximated by $\sigma_1(\omega)=1/\rho$, where $\rho$ is the dc resistivity. Eq.~(\ref{Eq:R}) then becomes the Hagen-Rubens relation \cite{Hagenrubens1903}
	\begin{equation}\label{EqSup:HagenRubens}
		R(\omega)\approx1-\sqrt{{8\epsilon_0\omega\rho}}.
	\end{equation}
A fit of this formula to the low-frequency reflectivity, shown in the inset of
Fig.~\ref{FigSup:Reflectivity} as dashed lines, yields the temperature-dependent
resistivity $\rho(T)$ shown in Fig.~1 of the main text.
%\ref{Fig:OpticalConductivity}
%
\section{Drude-Lorentz analysis, memory function}
\label{SM-II}
\begin{figure}[b]
\includegraphics[width=0.85\columnwidth]{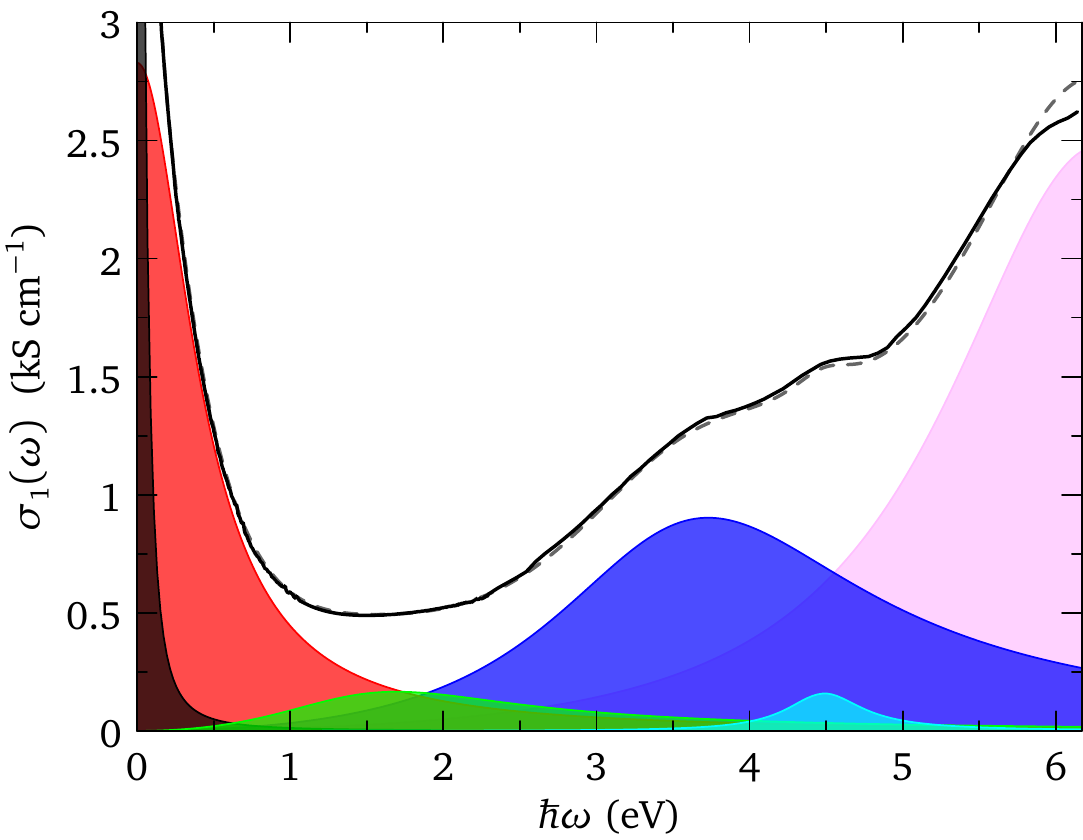}
\caption{\label{FigSup:SigmaLorentz}
Real part of the optical conductivity at 290~K, calculated from the
Drude-Lorentz model given by Eq.~(\ref{EqSup:DLeps}) and
Table~\ref{TabSup:DLparams} (dashed line). The shaded curves show the
contributions of the six oscillators, with the two Drude oscillators in black
and red. The solid line is the conductivity obtained directly from the
reflectivity and ellipsometry data using Kramers-Kronig relations.
}
\end{figure}

The in-plane reflectivity and ellipsometry data are fitted simultaneously using
the standard Drude-Lorentz model. The low-energy response below $\sim1$~eV is
well described at all temperatures by the superposition of two Drude
(zero-frequency) oscillators. Above 1~eV, the data are fitted with four
additional Lorentz oscillators. The high-frequency limit is represented by the
constant $\epsilon_{\infty}$, leading to the following model for
$\epsilon(\omega)$:
	\begin{equation}\label{EqSup:DLeps}
		\epsilon(\omega)=\epsilon_{\infty}+\sum_{j=1}^6\frac{\omega_{pj}^2}
		{\omega_j^2-\omega(\omega+i\gamma_j)}.
	\end{equation}
The parameters of the model fitted to the room-temperature data are given in
Table~\ref{TabSup:DLparams}. The real part of the conductivity calculated using
this model is displayed in Fig.~\ref{FigSup:SigmaLorentz}, and compared with the
conductivity determined directly by Kramers-Kronig transform of the
reflectivity. The match is excellent up to 5.8~eV.
\begin{table}
\renewcommand{\arraystretch}{1.5}
\caption{\label{TabSup:DLparams}
Parameters of the Drude-Lorentz oscillators for the in-plane optical response of
\sro{} at 290~K. All numbers are in eV. The high-frequency dielectric constant
is $\epsilon_{\infty}=2.3$. The intraband plasma frequency obtained from the
two Drude oscillators is $\hbar\omega_p=3.4$~eV.
}
\begin{tabular*}{0.9\columnwidth}{c@{\extracolsep{\fill}}cccccc}
\hline\hline
$j$ & 1 & 2 & 3 & 4 & 5 & 6 \\
\hline
$\hbar\omega_j$    & 0.000 & 0.000 & 1.693 & 3.730 & 4.489 & 6.298 \\
$\hbar\gamma_j$    & 0.038 & 0.432 & 1.952 & 2.533 & 0.602 & 2.527 \\
$\hbar\omega_{pj}$ & 1.568 & 3.014 & 1.552 & 4.123 & 0.843 & 6.826 \\
\hline\hline
\end{tabular*}
\end{table}

The Drude-Lorentz analysis allows us to distinguish the contribution of the
mobile charge carriers from that of the bound charges. At all temperatures
considered, the analysis yields two zero-frequency modes, as well as
finite-energy contributions that are all above 1~eV. The two Drude modes mimic
the ``foot'' structure of the low-energy Fermi-liquid response described in the
main text. We ascribe the two Drude oscillators to the intraband response of
mobile carriers, and all finite-energy modes to the bound charges. The spectral
weight of the charge carriers is, in units of $\epsilon_0$,
	\begin{equation}\label{EqSup:wptot}
		\omega_p^2\equiv\sum_{\hbar\omega_j=0}\omega_{pj}^2.
	\end{equation}
This gives a spectral weight that is almost temperature independent,
corresponding to $\hbar\omega_p=3.3$~eV at low temperature and 3.4~eV at room
temperature. Note that the spectral weight of the sharpest Drude mode is smaller,
and decreases from $\hbar\omega_{p1}=1.8$~eV at low temperature to 1.6~eV at
room temperature.

\begin{figure}[b]
\includegraphics[width=0.85\columnwidth]{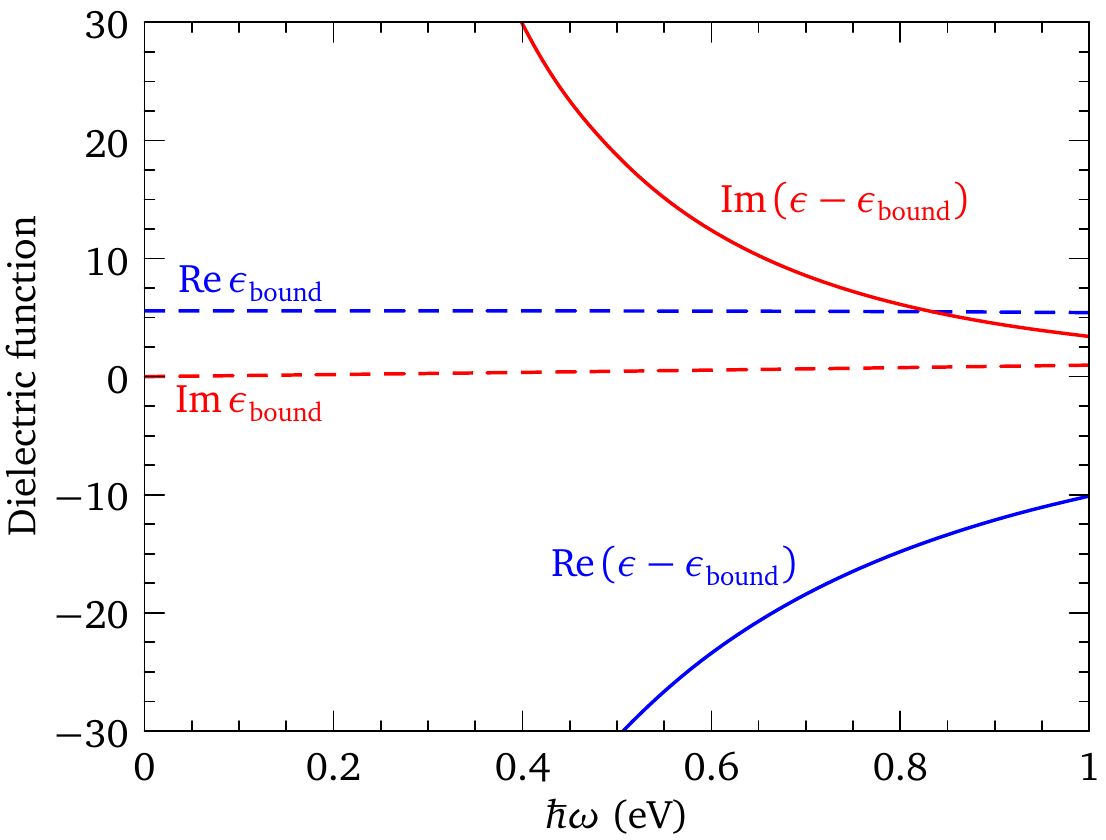}
\caption{\label{FigSup:DielectricFunctionSeparated}
Mobile- and bound-charge contributions to the dielectric function at 290~K. The
negligible frequency dependence of the bound-charge response indicates a clear
separation between the intraband and interband responses below 1~eV.
}
\end{figure}

In the energy range $\hbar\omega<1$~eV of interest for the present study, the
interband transitions do not contribute to the frequency dependence of the
dielectric function, as can be seen in
Fig.~\ref{FigSup:DielectricFunctionSeparated}. The bound-charge dielectric
function defined as
	\begin{align}\label{EqSup:epsb}
		\epsilon_{\mathrm{bound}}(\omega)&=\epsilon_{\infty}+
		\sum_{\hbar\omega_j>0}\frac{\omega_{pj}^2}
		{\omega_j^2-\omega(\omega+i\gamma_j)}\\
		&\approx\epsilon_{\infty}+\sum_{\hbar\omega_j>0}\frac{\omega_{pj}^2}
		{\omega_j^2}
	\end{align}
is practically constant over this energy range, and equal at 290~K to
$\epsilon_{\mathrm{bound}}(0)=5.6$.

\begin{figure}
\includegraphics[width=0.85\columnwidth]{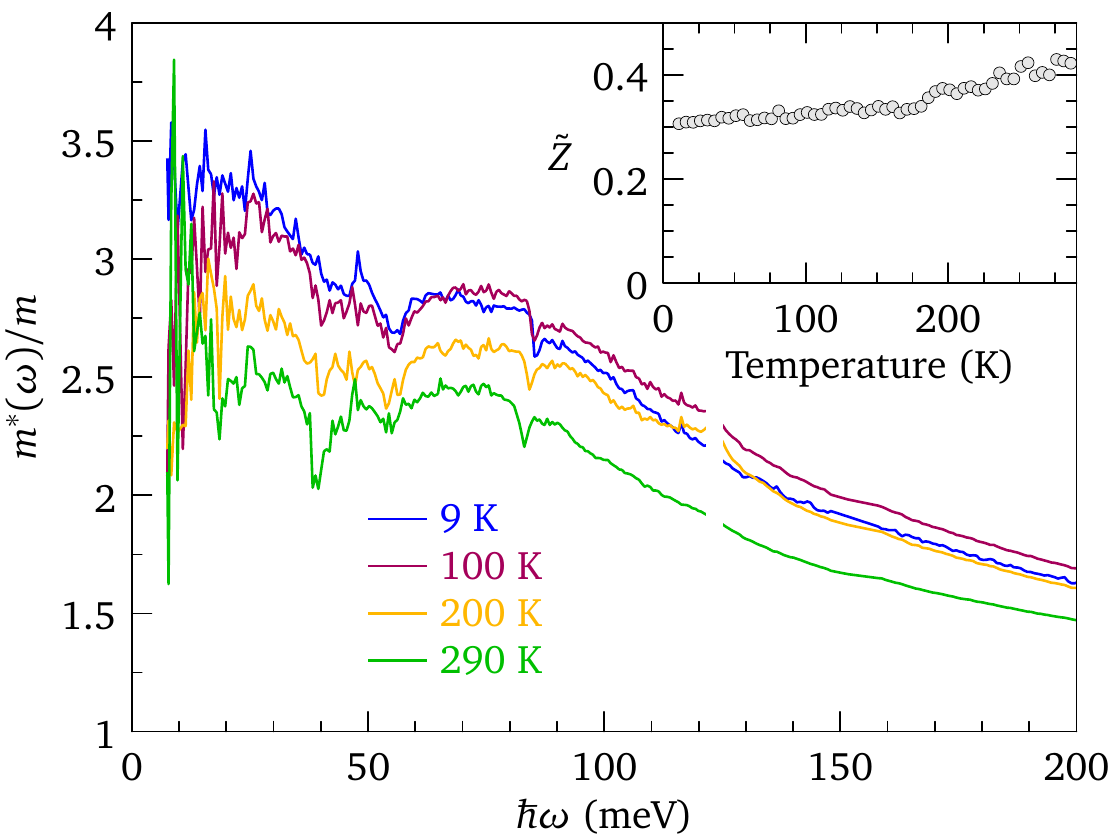}
\caption{\label{Fig:Sup_Mstar}
Frequency-dependent mass enhancement $m^*(\omega)/m=1+M_1(\omega)/\omega$
for some selected temperatures, and temperature dependence of
$\tilde{Z}=m/m^*(\omega)$ extrapolated to zero frequency. The white space close to 130 meV indicates the region where two data sets have been linked.
}
\end{figure}

Knowing the intraband spectral weight $\epsilon_0\omega_p^2$, we may invert
Eq.~(1) of the main text, and express the memory function, or %\ref{Eq:SigmaMemory}
optical self-energy $M(\omega)=M_1(\omega)+iM_2(\omega)$, in terms of the
optical conductivity:
	\begin{align}
		 1+\frac{M_1(\omega)}{\omega}&\equiv\frac{m^*(\omega)}{m}
		 =\mathrm{Im}\,\left[\frac{-\epsilon_0\omega_p^2 }{\omega\sigma(\omega)}\right]\\
		 M_2(\omega)&\equiv\frac{1}{\tau(\omega)}
		 =\mathrm{Re}\,\left[\frac{\epsilon_0\omega_p^2 }{\sigma(\omega)}\right].
	\end{align}
The memory function is displayed for selected temperatures in
Fig.~2 of the main text, and the mass enhancement factor %\ref{Fig:MemoryFunction}
in Fig.~\ref{Fig:Sup_Mstar}. It presents an overall behavior
determined by the charge-carrier dynamics, as well as sharp superimposed
Fano-like structures corresponding to dipole active optical phonons. We do not
subtract phonons, which would require a specific modeling and introduce unwanted
ambiguities.
\section{Fermi-liquid analysis}
\label{SM-III}
The universal characteristics of the optical response in a Fermi liquid have
been described in Ref.~\onlinecite{berthod_non-drude_2013}. For a local Fermi
liquid, i.e., with a momentum-independent self-energy, the low-energy optical
conductivity normalized to the dc conductivity is a universal function of the
two variables $\hbar\omega/(2\pi\kB T)$ and $\omega\tau_{\text{qp}}$, where
$\tau_{\text{qp}}$ is the quasiparticle life-time on the Fermi surface. This
life-time diverges as $1/T^2$ with decreasing temperature in a Fermi liquid. In
Ref.~\onlinecite{berthod_non-drude_2013}, it was parametrized by a temperature
scale $T_0$, such that $\hbar/\tau_{\text{qp}}$ equals $2\pi\kB T$ at this
temperature: $\hbar/\tau_{\text{qp}}=2\pi(\kB T)^2/(\kB T_0)$. The parameter
$T_0$ completely determines the universal behavior of
$\sigma(\omega)/\sigma_{\text{dc}}$. The non-universal dc conductivity depends
on the quasiparticle residue $Z$, and on the transport function $\Phi(0)$
proportional to the average squared velocity on the Fermi surface. In the dc
conductivity, these two quantities enter as the product $Z\Phi(0)$. This product
also coincides with the weight of the Drude peak. The Fermi-liquid model can be
generalized to include impurity scattering in the form of a constant scattering
rate $\Gamma$, entering the optical conductivity as the product $Z\Gamma$. Thus
the low-energy optical conductivity of a local Fermi liquid can be represented
by the three parameters $T_0$, $Z\Phi(0)$, and $Z\Gamma$. The temperature scale
$T_0$ is a relatively large scale, typically ten times larger than the scale
$\TFL$, below which the universal Fermi-liquid behavior is generally observed
\cite{berthod_non-drude_2013}.

\begin{figure}[b]
\includegraphics[width=0.9\columnwidth]{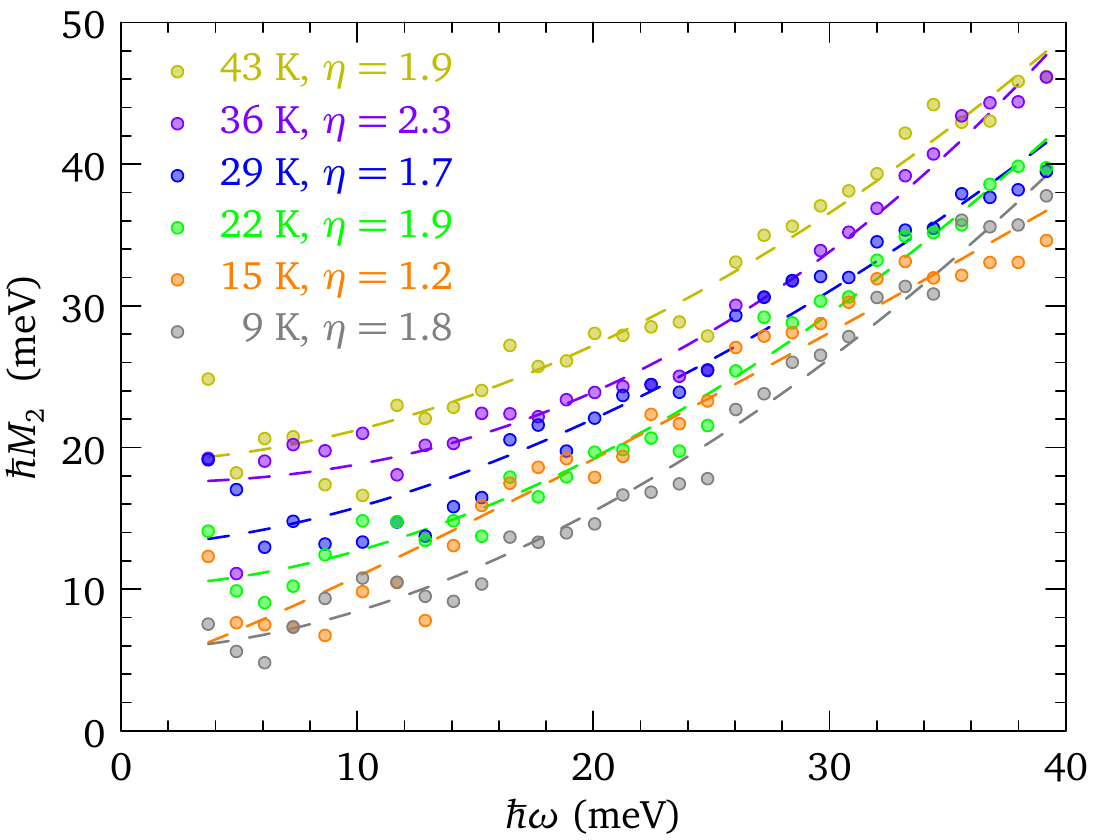}
\caption{\label{fig:scatteringsquared}
Imaginary part of the memory function binned in 2~meV intervals (points) for selected temperatures. The data is fitted below 40~meV using the power law $M_2(\omega) = M_2(0)+A\omega^{\eta}$ (dashed lines). The exponent $\eta$ is indicated for each 
temperature.}
\end{figure}

In the present study, the experimental intraband response is represented by the
complex memory function $M=M_1+iM_2$,
$\sigma(\omega)=i\epsilon_0\omega_p^2/[\omega+M(\omega)]$, and the reference
spectral weight $\epsilon_0\omega_p^2$ is taken as the intraband weight,
determined as explained in Sec.~\ref{SM-II}. In the thermal regime,
$\hbar\omega\sim 2\pi\kB T$, the Fermi-liquid memory function has the form
	\begin{multline}\label{eq:FLM}
		M(\omega)\approx\left(\frac{1}{\tilde{Z}}-1\right)\omega\\
		+\frac{2i}{\hbar\tilde{Z}}
		\left(\frac{(\hbar\omega)^2+(2\pi\kB T)^2}{3\pi\kB T_0}+Z\Gamma\right).
	\end{multline}
$\tilde{Z}=Z\Phi(0)/(\epsilon_0\omega_p^2)$ is the ratio of the Drude weight to
the reference weight. In the Fermi-liquid regime, a plot of $\hbar M_2$ as a
function of $\xi^2=(\hbar\omega)^2+(2\pi\kB T)^2$ yields a straight line with a
slope $2/(3\pi\kB T_0\tilde{Z})$, and an intercept $2Z\Gamma/\tilde{Z}$. For
Sr$_2$RuO$_4$, this scaling behavior is observed for temperatures below
$\sim 40$~K and frequencies below $\sim 36$~meV, as shown in
Fig.~3 of the main text. %\ref{Fig:ScalingRelaxationRate} 
Fitting $M_2(\omega)$ using a power law in that frequency and temperature range (Fig.~\ref{fig:scatteringsquared}) indicates a clear curvature $\omega^{\eta}$ with
$\eta=1.7$--$2.3$, with one outlier ($T=15$~K), where the least-squares fit gives $\eta=1.15$.

\begin{figure}[t]
\includegraphics[width=0.9\columnwidth]{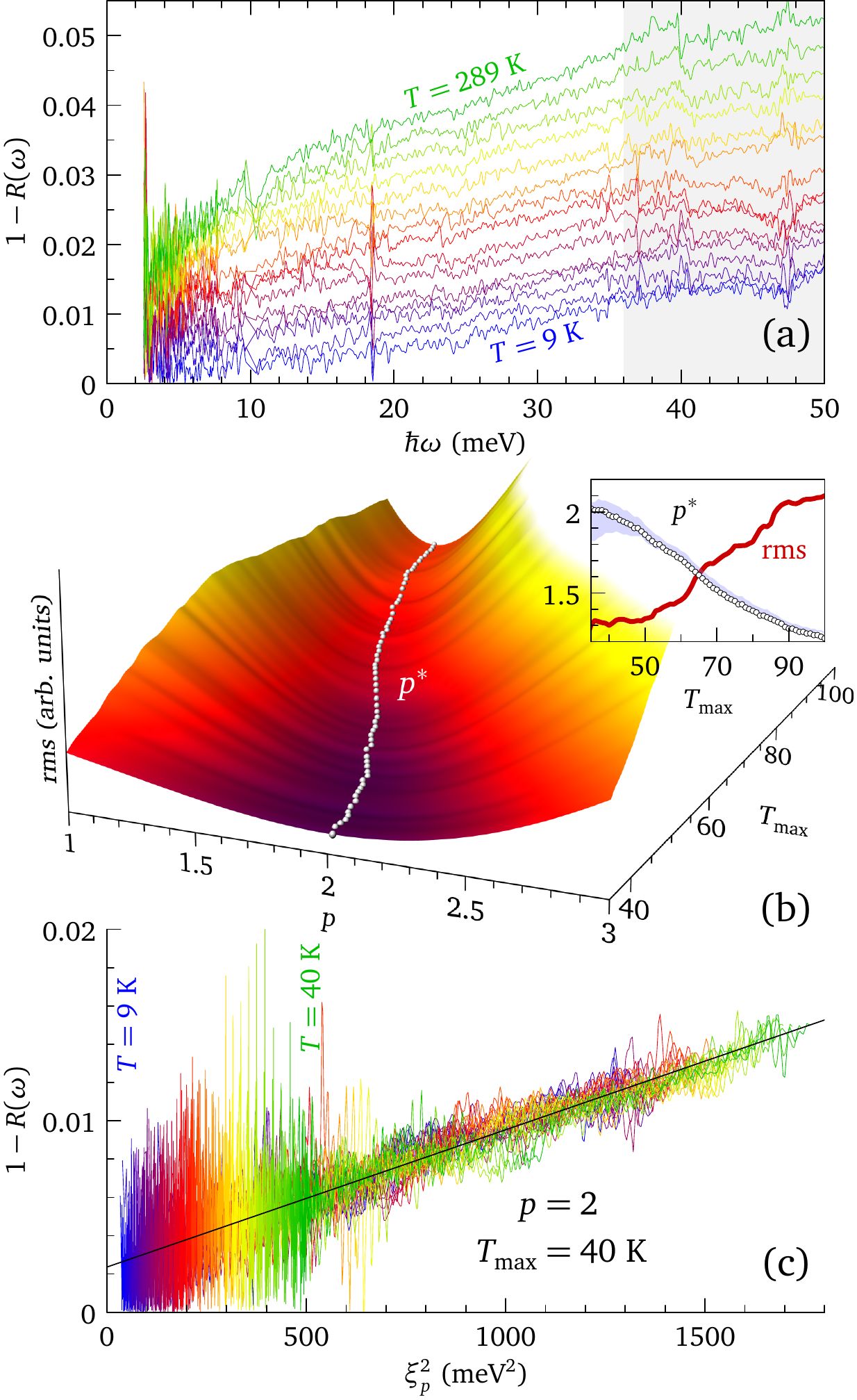}
\caption{\label{fig:scaling-reflectivity}
(a) Reflectivity data plotted as $1-R(\omega)$ versus $\hbar\omega$ for
$9~\text{K}\leqslant T\leqslant 290$~K (one curve every 20~K). The data in the shaded
region ($\hbar\omega>36$~meV) is ignored in the
Fermi-liquid scaling analysis. (b) Root-mean square deviation of the
reflectivity for $T\leqslant T_{\max}$ from a linear dependence in $\xi_p^2$, as
a function of $p$ and $T_{\max}$. The inset shows the value $p^*$ (dots) and the
rms (solid, arbitrary units) at the minimum, versus $T_{\max}$. The shaded
region shows how $p^*$ changes if the range of frequencies retained in the
analysis is varied by $\pm5$~meV. (c) Collapse of the reflectivity data for
$T\leqslant40$~K. Blue corresponds to $T=9$~K and green to $T=40$~K. The linear
correlation coefficient is 0.9.
}
\end{figure}

The frequency and temperature scaling properties of the optical relaxation rate in the thermal regime, $\hbar\omega\sim2\pi\kB T$, 
can also be observed in the bare reflectivity data at normal incidence. We substitute the definition of the memory function in Eq.~(\ref{Eq:R}), and expand in powers of $M_2/(\omega+M_1)$:
\begin{multline}\label{eq:RFLM}
		1-R(\omega)\approx
		\sqrt{\frac{8\epsilon_0\omega}{Z\Phi(0)}\left(M_2+\sqrt{(\omega+M_1)^2+M_2^2}\right)} =\\
		2\sqrt{\frac{\epsilon_0}{Z\Phi(0)}} \frac{M_2}{\omega+M_1} +O\left(\left[ \frac{M_2}{\omega+M_1}\right]^2\right).
\end{multline}
Substituting Eq.~(\ref{eq:FLM}) for $M(\omega)$ we find that, in the thermal regime, $1-R(\omega)$ behaves like $M_2(\omega)$:
	\begin{multline}
		1-R(\omega)=
		\frac{4}{\hbar}\sqrt{\frac{\epsilon_0}{Z\Phi(0)}}
		\left[\frac{(\hbar\omega)^2+(2\pi\kB T)^2}{3\pi\kB T_0}+Z\Gamma\right].
	\end{multline}
In order to test the presence of this scaling in the \sro{} reflectivity, we
proceed like for the imaginary part of the memory function in the main text: for
all temperatures $T\leqslant T_{\max}$ and all frequencies $\hbar\omega\leqslant
36$~meV, we plot
$1-R(\omega)$ as a function of $\xi_p^2=(\hbar\omega)^2+(p\pi\kB T)^2$, and we
determine the value of $p$ which minimizes the rms deviation of the data from a
straight line. The result of this analysis is shown in
Fig.~\ref{fig:scaling-reflectivity}. It leads to the same conclusion as the
analysis performed on the memory function: the Fermi-liquid universal behavior
with $p=2$ is only seen below $T\sim 40$~K. We note that the optimal value of
$p$ depends to some extent on the window of frequencies considered. Extending
the window up and down by 5~meV leads to changes in the curve $p^*(T_{\max})$ as
indicated in the inset of Fig.~\ref{fig:scaling-reflectivity}(b). With the
reflectivity data, which is not affected by uncertainties of the Kramers-Kronig
transform, the curve is less sensitive to the frequency window than with the
memory-function data of Fig.~3 in the main text.%\ref{Fig:ScalingRelaxationRate}

\begin{figure}[b]
\includegraphics[width=0.9\columnwidth]{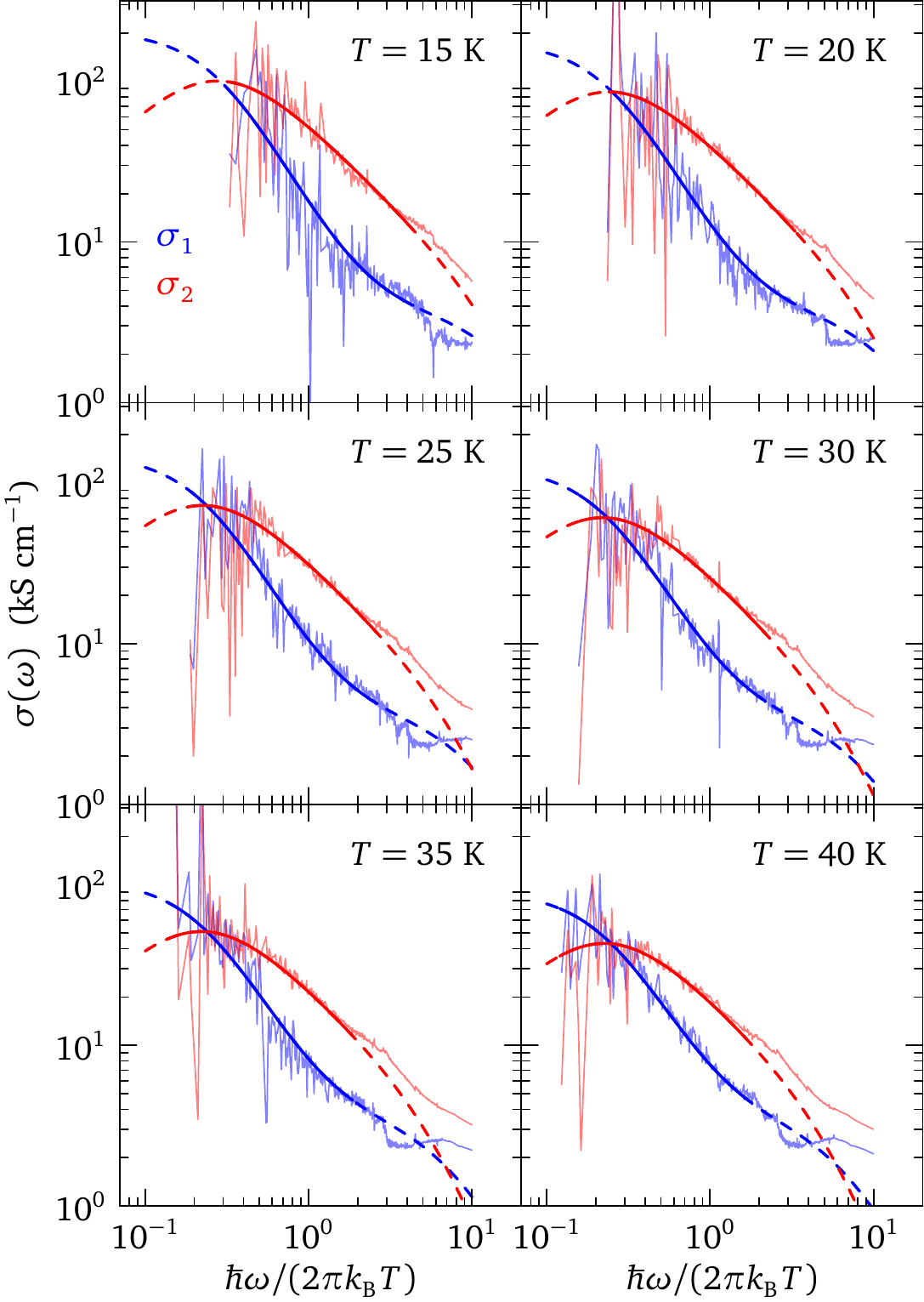}
\caption{\label{fig:conductivity-FL}
Optical conductivity of \sro{} in the Fermi-liquid regime. The thin lines show
the real ($\sigma_1$, blue) and imaginary parts ($\sigma_2$, red) of the
measured conductivity. The thick lines show the fit to the Fermi-liquid model of
Ref.~\onlinecite{berthod_non-drude_2013}, with the parameters $T_0=316$~K,
$\hbar^2Z\Phi(0)=3.5\epsilon_0\text{eV}^2$, and $Z\Gamma=0.9$~meV. The lines
are solid in the frequency range below
$\hbar\omega\leqslant36~\text{meV}$ considered for the
fit, and dashed outside this range.
}
\end{figure}

In the frequency and temperature domain where the above analysis points to
Fermi-liquid behavior with $p=2$, the real and imaginary parts of the
Sr$_2$RuO$_4$ optical conductivity display a characteristic change of curvature
around $\hbar\omega=2\pi\kB T$. This can be seen in
Fig.~\ref{fig:conductivity-FL}: the low-frequency Drude behavior of $\sigma_1$
changes to a weaker frequency dependence for $\hbar\omega>2\pi\kB T$. A similar
change of slope was identified in Ref.~\onlinecite{berthod_non-drude_2013} to be
a signature of the universal Fermi-liquid response. Adjusting the local-Fermi
liquid model to the data in the domain $\hbar\omega\leqslant36~\text{meV}$ and $T\leqslant40$~K,
we find that it provides a very good fit of the measured conductivity (see
Fig.~\ref{fig:conductivity-FL}). This should be regarded as an effective
one-band description of the three-band response of Sr$_2$RuO$_4$. The effective
model parameters resulting from the fit are $T_0=316$~K,
$\hbar^2Z\Phi(0)=3.5\epsilon_0\text{eV}^2$, and $Z\Gamma=0.9$~meV. The Drude
weight $Z\Phi(0)$ corresponds to a plasma frequency $\approx 1.87$~eV, in
good agreement with the value of $\hbar\omega_{p1}$ obtained from the Drude-Lorentz analysis at
$T=9$~K (Sec.~\ref{SM-II}).
\section{\textit{Ab initio} calculations}
\label{SM-IV}
\subsection{Method}
We calculated the optical conductivity within an \textit{ab-initio} framework that
combines density-functional theory (DFT) in the local-density
approximation (LDA) as implemented in Wien2k \cite{Wien2k}, with the
many-body dynamical mean-field theory (DMFT) \cite{Georges1996}. The
framework is described in Ref.~\onlinecite{aichhorn_2009}.

The optical conductivity in DFT+DMFT is expressed as 
	\begin{multline}\label{eq:opt_dmft}
		\sigma(\omega)=\frac{2\pi e^2}{V}\sum_{\vec{k}}\int_{-\infty}^{\infty}
		d\varepsilon\,\frac{f(\varepsilon)-f(\varepsilon+\hbar\omega)}{\omega}\\
		\times\mathrm{Tr}\,v_{\vec{k}}^x A_{\vec{k}}^{\phantom{x}}(\varepsilon)
		v_{\vec{k}}^x A_{\vec{k}}^{\phantom{x}}(\varepsilon+\hbar\omega).
	\end{multline}
$V$ is a normalization volume, $f(\varepsilon)$ is the Fermi function, $\vec{v}_{\vec{k}}$
and $A_{\vec{k}}(\varepsilon)$ are the band velocities and the spectral functions, respectively,
both
evaluated at wave-vector $\vec{k}$. $v_{\vec{k}}^x$ and $A_{\vec{k}}$
are matrices in the band indices. The velocities are obtained from DFT
as described in Refs.~\onlinecite{claudia2006, Wien2k}, and the
spectral functions
are calculated as described in a previous work \cite{aichhorn_2009,mravlje_2011},
using the same interaction parameters.
\subsection{Insignificance of the band velocities for the intraband conductivity}
\begin{figure}[b]
\includegraphics[width=0.8\columnwidth]{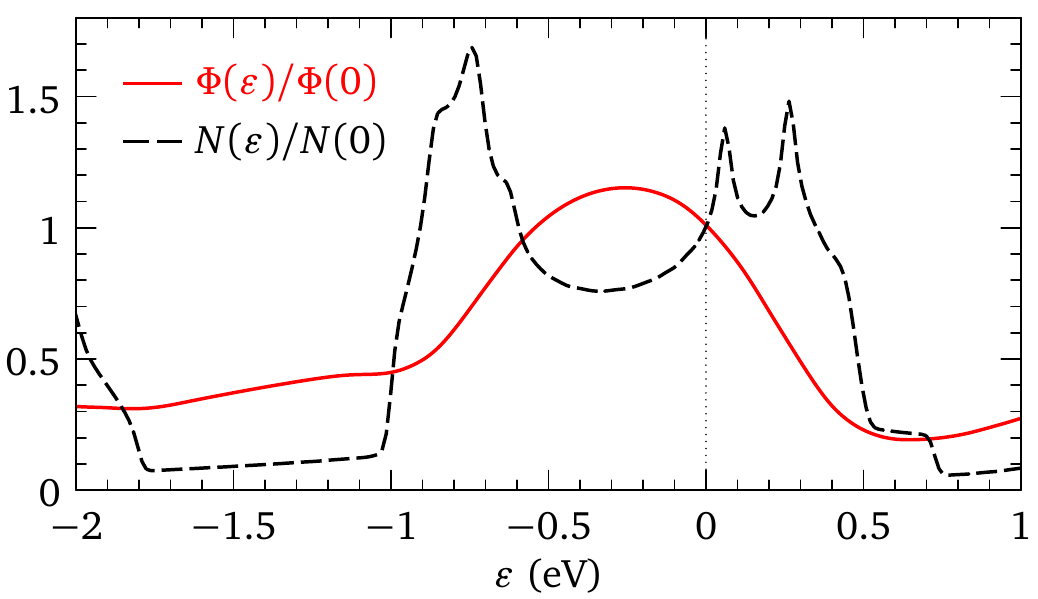}
\caption{\label{fig:dos}
\sro{} in-plane transport function $\Phi(\varepsilon)$, and
DOS $N(\varepsilon)$, normalized to their Fermi-level values.
$\Phi(0)=\epsilon_0\omega_p^2=0.38\times 10^{21}~(\Omega\mathrm{ms})^{-1}$
($\hbar\omega_p=4.3$~eV), in reasonable agreement with previous results
\cite{oguchi1995, singh_relationship_1995}.
}
\end{figure}
We first address the question whether the increase of optical conductivity
reported in the main text for energies above 0.1~eV, with respect to the universal
Fermi-liquid (FL) behavior, could be a consequence
of the bare electronic dispersions. The possible effects of the band structure
must be considered, because the density of states (DOS) of \sro{} has a rich structure
at low energy, in particular, a Van Hove singularity
about 70~meV above the Fermi level (see Fig.~\ref{fig:dos}). The quantity that is relevant for
the optical properties is not the DOS, though, but the
transport function
	\begin{equation}
		\Phi(\varepsilon)=\frac{2e^2}{V}\sum_{\vec{k}}
		\mathrm{Tr}\,\left(v_{\vec{k}}^x v_{\vec{k}}^x\right)\,
		\delta(\varepsilon-\varepsilon_{\vec{k}}).
	\end{equation}
As can be seen in Fig.~\ref{fig:dos}, the transport function is much
more smooth than the DOS, and presents no significant feature at the energy of the
Van Hove singularity. This can be understood, since the band velocity
vanishes at the Van Hove points.

\begin{figure}
\includegraphics[width=0.9\columnwidth]{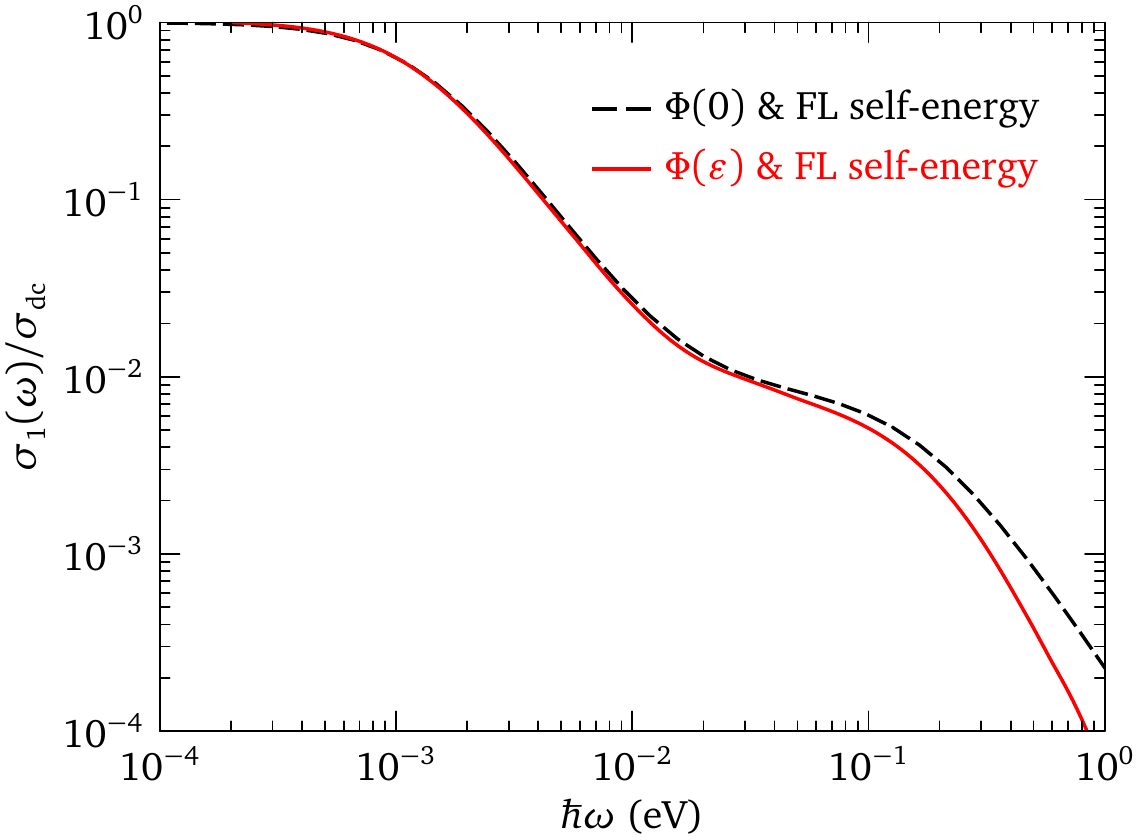}
\caption{\label{fig:trial_FL}
Role of the energy dependence of the transport function $\Phi(\varepsilon)$. The solid
line is the real part of the conductivity obtained by using the same Fermi-liquid self-energy,
Eq.~(\ref{eq:self_fl}), for the three $t_{2g}$ bands. The dashed line is the universal
FL result,
where the energy dependence of $\Phi(\varepsilon)$ is neglected.}
\end{figure}

The absence of structure in $\Phi(\varepsilon)$
suggests that the excess conductivity above 0.1~eV is not due to the band
structure. To be more quantitative, we have constructed a simplified spectral function,
taking for all orbitals the FL self-energy Ansatz \cite{berthod_non-drude_2013}
	\begin{equation}\label{eq:self_fl}
		\Sigma(\varepsilon)=\left(1-\frac{1}{Z}\right)\varepsilon
		-\frac{i}{Z\pi\kB T_0}\left[\varepsilon^2+ (\pi\kB T)^2\right],
	\end{equation}
instead of $\Sigma_{\vec{k}}(\varepsilon)$. We used
parameters that fit the self-energy of the {\it xy} band at low energy for
$T=29$~K, namely, $1/Z=5.12$ and $T_0=400$~K. This model reproduces the
universal FL result if $\Phi(\varepsilon)$ is approximated by
$\Phi(0)$. In Fig.~\ref{fig:trial_FL}, we compare the real part of
the conductivity obtained using the full $\varepsilon$-dependence of
$\Phi(\varepsilon)$ and the FL result. The energy dependence of the
transport function does produce a deviation from the FL curve, but in the
direction of a \emph{smaller} conductivity, which is opposite to the
experimentally observed deviation.
\subsection{DFT+DMFT self-energies}
\begin{figure}[b]
\includegraphics[width=0.85\columnwidth]{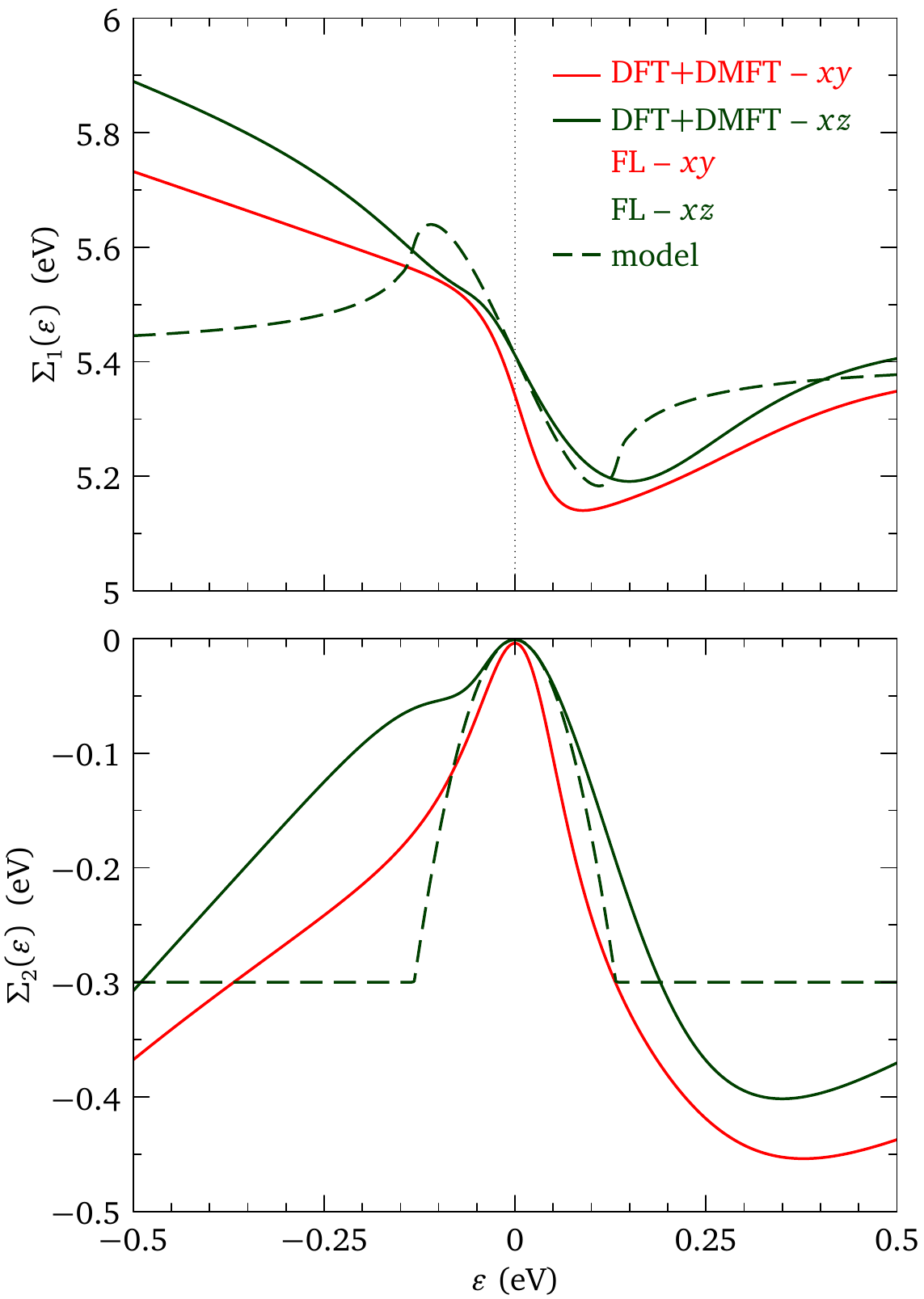}
\caption{\label{fig:self}
Local self-energies for the {\it xy} and {\it xz} bands at $T =$~29~K. The real part (top) is linear at
low energy, and the imaginary part (bottom) is quadratic, as shown by the dotted lines.
The dashed lines show a model with bounded quadratic imaginary part, and the
corresponding real part.
}
\end{figure}

The local DMFT self-energies for the {\it xy} and {\it xz/yz} orbitals are shown in Fig.~\ref{fig:self}.
At low energy, both follow the FL behavior, with a parabolic dependence of the imaginary part and a linear dependence of the real part. As discussed earlier \cite{mravlje_2011}, the {\it xy} orbital is more correlated, with steeper real part and stronger curvature of
the imaginary part. The self-energies start to deviate from strict FL behavior already at low energy ($\sim0.03$~eV), and become markedly electron-hole asymmetric.
As seen more clearly
in the real part, differences with FL dependence become substantial above 0.05~eV,
especially on the electron side, where a drastic change of slope is observed. 

Fig.~\ref{fig:self} also displays a simple model for the self-energy, with an imaginary part which is purely quadratic 
up to a cutoff frequency $\varepsilon_c=0.125$~eV, and constant with a value $\gamma_c=-0.3$~eV above this cutoff.
The corresponding real-part, displayed in Fig.~\ref{fig:self}, is obtained by Kramers-Kronig as: 
\begin{equation}
\mathrm{Re}\,\Sigma(\varepsilon)=\frac{\gamma_c}{\pi}\left(\frac{\varepsilon^2}{\varepsilon_c^2}-1\right)
\ln\left|\frac{\varepsilon-\varepsilon_c}{\varepsilon+\varepsilon_c}\right|+
\frac{2\gamma_c}{\pi}\frac{\varepsilon}{\varepsilon_c}.
\end{equation}
This model reproduces the main qualitative aspects of the data, especially the strong feature in the real part. 

\subsection{Analysis of the optical response}
\begin{figure}[b]
\includegraphics[width=\columnwidth]{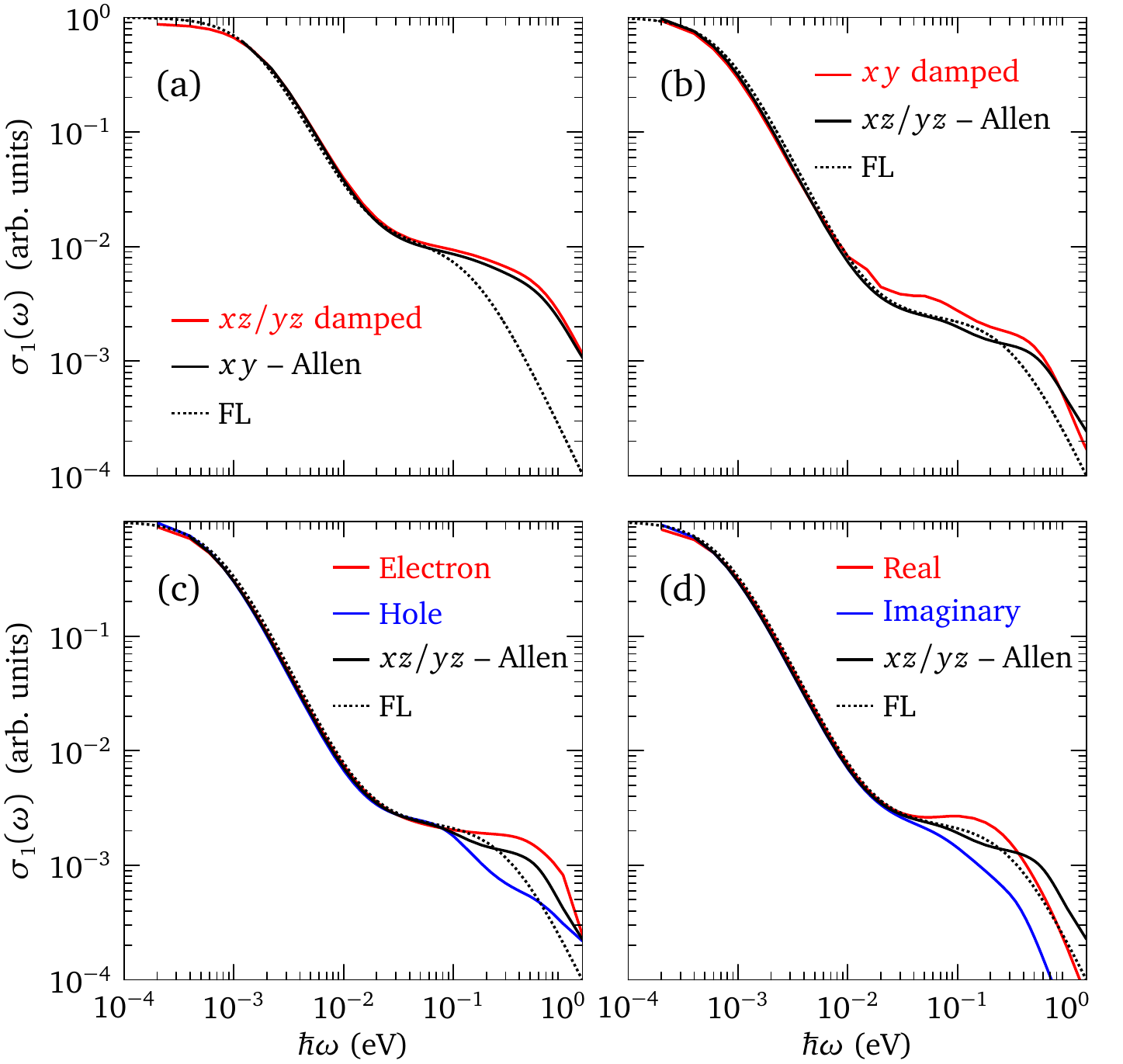}
\caption{\label{fig:surgery}
(a) Optical conductivity obtained from Eq.~(\ref{eq:opt_dmft}), after suppressing
the contributions of the $xz/yz$ orbitals (red), compared with the result of
Allen formula (\ref{eq:sigma_Sigma}) evaluated with the self-energy of the {\it xy} orbital (black), and with
the universal FL curve (dotted). (b) Same as (a), with the roles of {\it xy} and $xz/yz$ exchanged.
(c) Contribution of the $xz/yz$ orbitals ({\it xy} damped), with modified
self-energies having either electron, or hole character (see text).
(d) Contribution of the $xz/yz$ orbitals ({\it xy} damped), with modified self-energies
having DMFT real part and FL imaginary part (Real) or FL real part, and DMFT imaginary
part (Imaginary).
}
\end{figure}

In order to clarify the origin of the optical conductivity departure from the FL prediction
above 0.1~eV, we have performed several numerical experiments. First, it is convenient
to look separately at the contribution of each orbital to the conductivity.
We have done this in two ways: (i) we have
evaluated Eq.~(\ref{eq:opt_dmft}) using spectral functions in which one
of the orbital was heavily damped, by adding a very large
imaginary part to its self-energy; (ii) we have calculated the optical conductivity using
the Allen formula \cite{Allen2004}
	\begin{equation}\label{eq:sigma_Sigma}
		\sigma(\omega)=\frac{i\Phi(0)}{\omega}\int_{-\infty}^{\infty}
		d\varepsilon\,\frac{f(\varepsilon)-f(\varepsilon+\hbar\omega)}
		{\hbar\omega+\Sigma^*(\varepsilon)-\Sigma(\varepsilon+\hbar\omega)}
	\end{equation}
and the local self-energy of each band independently.
The results are displayed in Fig.~\ref{fig:surgery}(a) and (b). For both orbitals,
a deviation from FL is seen above 0.05--0.1~eV, in the direction of an increased
conductivity. One also sees that the Allen formula describes the data reasonably well.

Next, we investigated which of the electron or hole part of the self-energy
plays the key role in the extra conductivity. For this purpose, we defined
particle-like orbital self-energies $\Sigma_p(\varepsilon)$, whose dependence at negative
energy is obtained by the reflection of the positive-energy dependence. For the
imaginary part, $\Sigma_{p2}(\varepsilon)=\Sigma_2(|\varepsilon|)$, and for the
real part, $\Sigma_{p1}(\varepsilon<0)=2\Sigma_1(0)-\Sigma_1(-\varepsilon)$,
$\Sigma_{p1}(\varepsilon>0)=\Sigma_1(\varepsilon)$. Similarly, hole-like self-energies
are constructed by a reflection of the negative-energy data to positive energy.
The resulting optical conductivities (contribution of the $xz/yz$ orbitals) are
presented in Fig.~\ref{fig:surgery}(c), from which it is evident that the electrons 
(particle-like states of positive energy), rather than the holes, give rise to the extra spectral weight.

Lastly, we investigated whether the extra conductivity must be ascribed to the
real, or to the imaginary part of the self-energy. For this purpose, we constructed
trial self-energies, by replacing the real or imaginary part of
the DMFT self-energies with their low-energy FL extrapolations
(in strong violation of Kramers-Kronig relations).
The result is shown in Fig.~\ref{fig:surgery}(d). One sees
that keeping just the DMFT imaginary parts leads to a pronounced downward
deviation from the FL result. Only when the nonlinear real parts are included
does a deviation in the upward direction appear. This deviation, however, quickly dies
off with such trial self-energies that violate the Kramers-Kronig relations.
The reason is the rapid increase of scattering, as the FL result has already
entered the dissipative regime, where
increasing $|\Sigma_2|$ diminishes the conductivity.

\begin{figure}
\includegraphics[width=0.9\columnwidth]{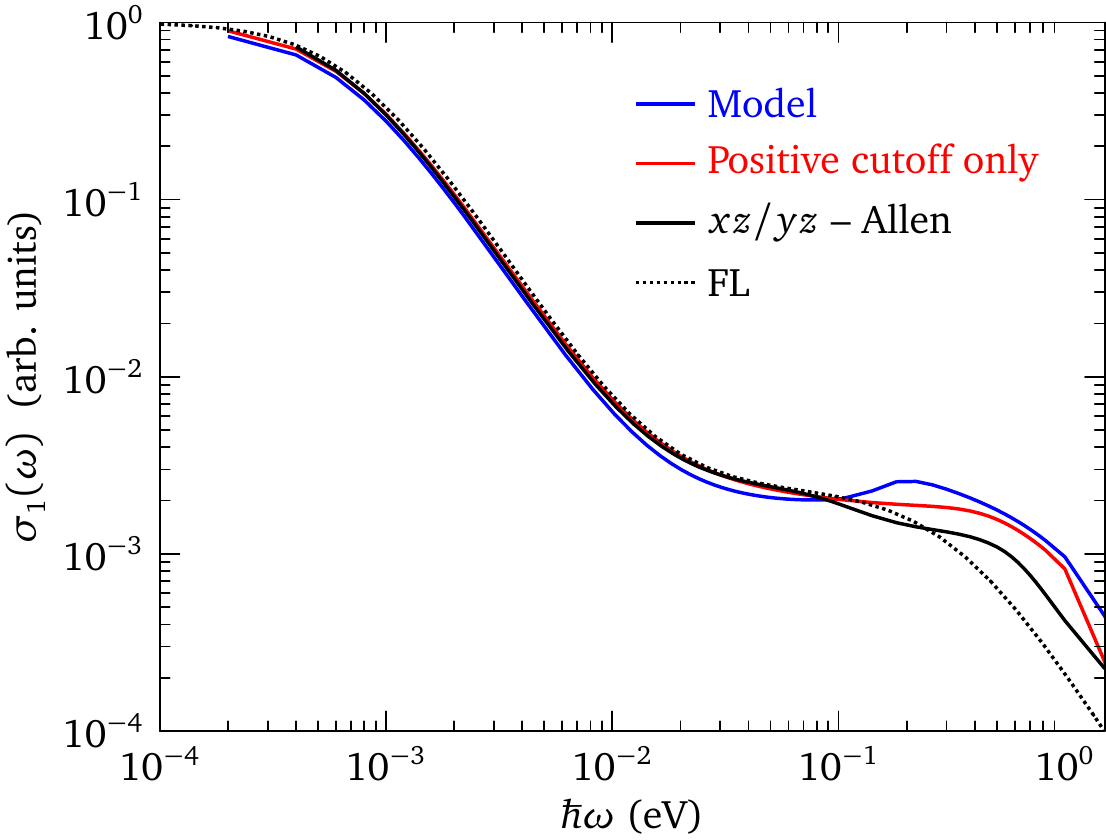}
\caption{\label{fig:model_cond}
Optical conductivities calculated using the Allen formula (\ref{eq:sigma_Sigma})
with the model self-energy of Fig.~\ref{fig:self}
(blue), and the self-energy of the $xz/yz$ orbitals (black), compared
with the FL result (dotted). The red curve is obtained by sending the
negative-energy cutoff of the model self-energy to $-\infty$.
}
\end{figure}

The deviation from FL dependence can be reproduced qualitatively by means of the model self-energy 
shown in Fig.~\ref{fig:self}. This self-energy has a quadratic imaginary part at low energy,
$-\Sigma_2(\varepsilon)\propto\varepsilon^2+(\pi\kB T)^2$ for $|\varepsilon|<\varepsilon_c$,
followed by a saturation for $|\varepsilon|>\varepsilon_c$. The real part
obtained by Kramers-Kronig displays sharp kinks at $\varepsilon_c$.
As shown in Fig.~\ref{fig:model_cond}, this very rough model reproduces
the data quite well. The description can be improved
if the parabolic dependence of the imaginary part is kept at negative
frequencies and saturation is only imposed on the positive side.

\subsection{Physical interpretation: the resilient quasiparticles} 

The essence of the departure from the FL form is thus related to the
sharp saturation of the scattering rate on the positive energy side,
and the related sharp feature in the real part at a scale of 0.1~eV. The
pronounced action takes place on the electron (positive energy) side. The
saturation of the scattering rate and the associated robust dispersing
resilient quasiparticle excitations were found in the context of the Hubbard
model \cite{Deng2013}. Strikingly, in \sro{} this
saturation occurs in a more pronounced way and leads to a strong 
change of slope of the real part of the self-energy. 

\begin{figure}[h!]
\includegraphics[width=\columnwidth]{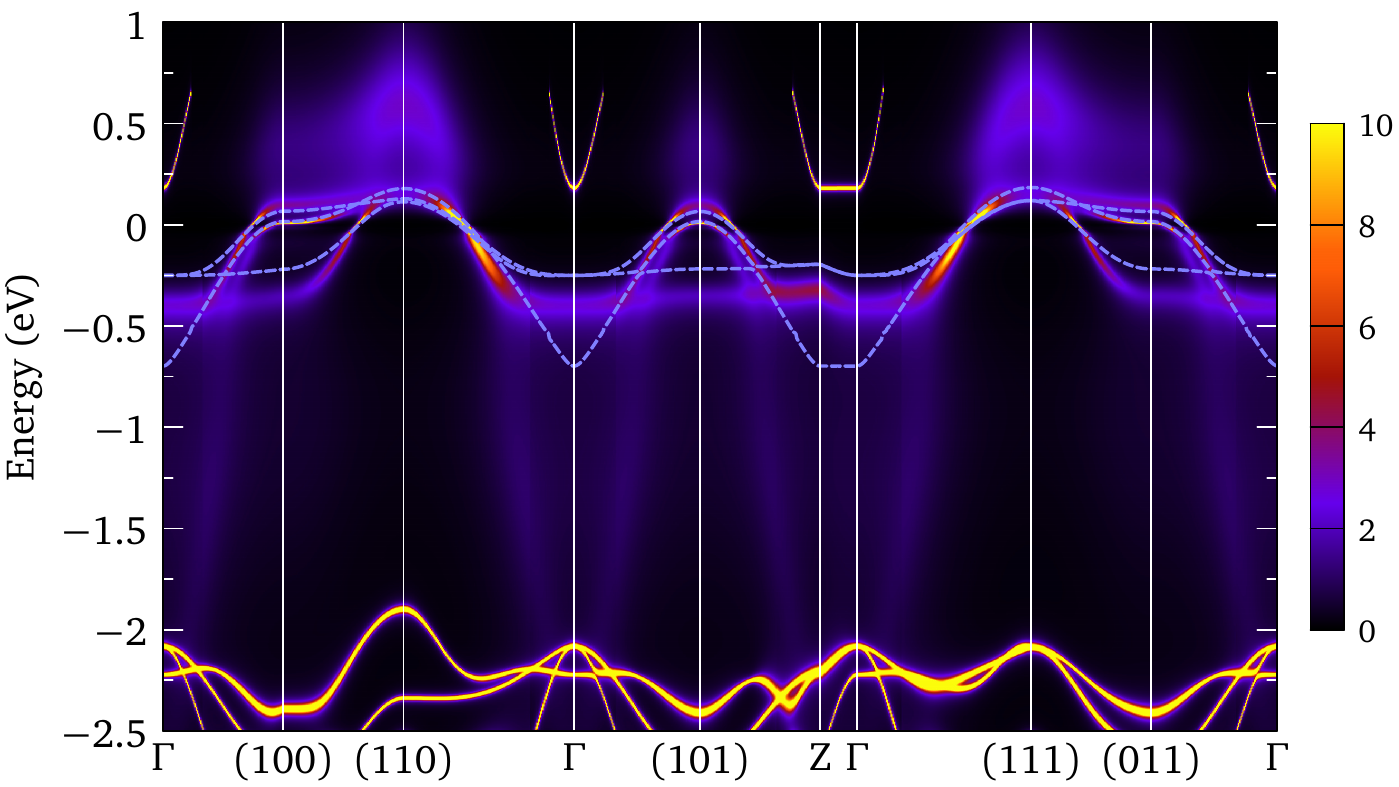} 
\caption{\label{fig:arpes}
Color map of the $\vec{k}$-resolved spectral function. The LDA $t_{2g}$
bands renormalized by a factor of 4 are shown with dashed lines.}
\end{figure}

The consequences of this change of slope can be most directly seen in
the color-map of the $\vec{k}$-resolved spectral function, that is presented
in Fig.~\ref{fig:arpes}. Superimposed are\linebreak

\begin{figure}[h]
%\vspace{-2em}
\includegraphics[width=\columnwidth]{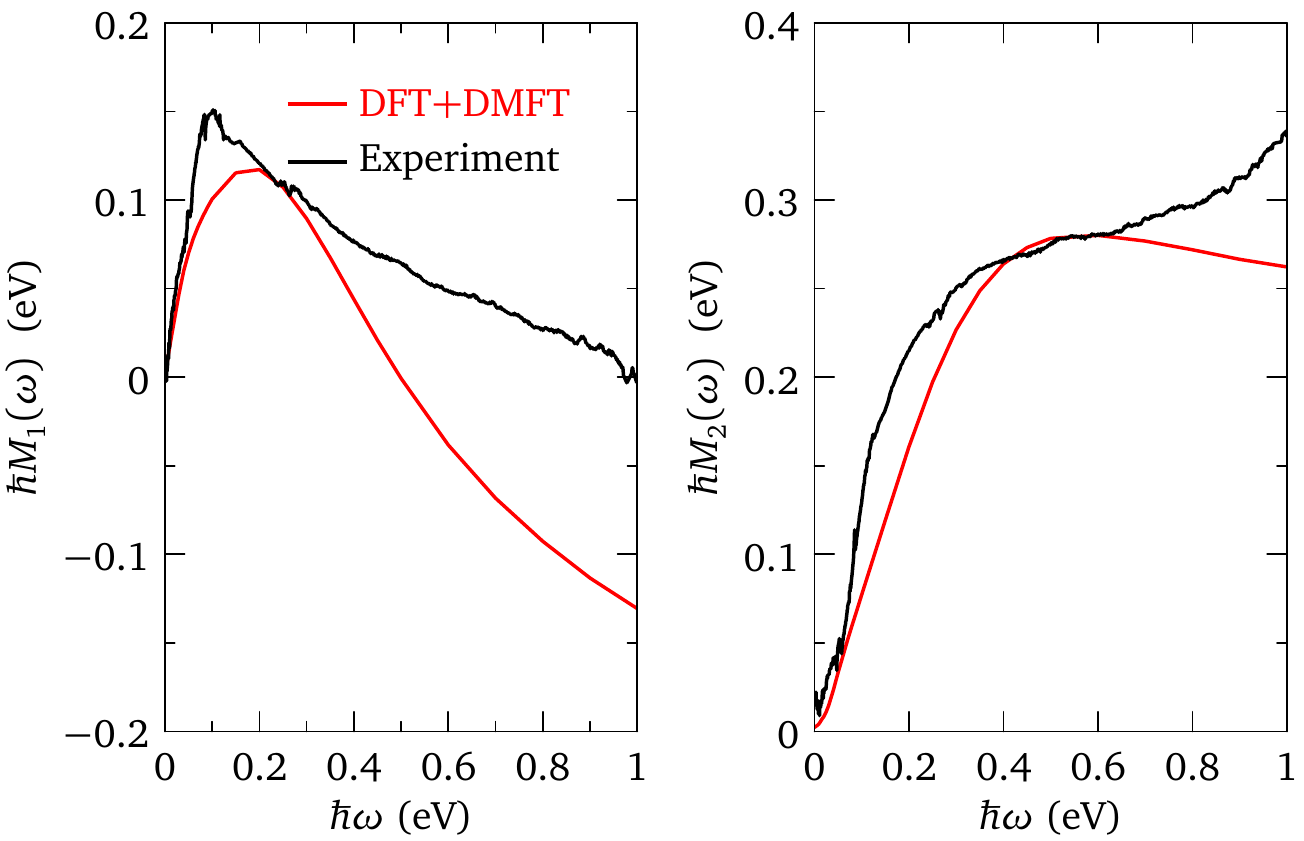}
\caption{\label{fig:memory}
Memory function; theory and experiment (both constructed using the experimentally determined $\hbar \omega_p=3.3$~eV).}
\end{figure}

\noindent also the LDA $t_{2g}$
bands that are renormalized by a factor of 4. Whereas at low energies
these renormalized bands describe the data reasonably well, above
$\sim 0.1$~eV (below $\sim -0.5$~eV) the dispersion abruptly increases,
giving rise to the pronounced inverted waterfall structure.

Above those energies, broad and strongly dispersing resilient
quasiparticle excitations appear very clearly in the $\vec{k}$-resolved
spectral function. The excess conductivity found in the experiment and
in our DFT+DMFT calculations is thus a consequence of highly
dispersive states that exist above the Fermi energy.
This is also reflected in the memory function, that is shown in
Fig.~\ref{fig:memory}. The change of slope that appears due to
the resilient quasiparticle states is seen clearly in the real part of
the theoretical and experimental memory function.


%%%%%%%%%%%%%%%%%%%%%%

%%%%%%%%%%%%%%
%\bibliography{SM_bib}%
%%%%%%%%%%%%%%

\end{document}